\documentclass[fleqn,usenatbib]{mnras}

\usepackage{mathptmx}

\usepackage[T1]{fontenc}
\usepackage{ae,aecompl}
\usepackage{color}

\usepackage{graphicx}	
\usepackage{amsmath}	
\usepackage{amssymb}	
\usepackage{txfonts}

\newcommand{\oc}{$O\!-\!C$}

\title[Accretion flow in  RW Tri]{ Structure of accretion flows in the nova-like cataclysmic variable~RW~Tri }

\author[Subebekova, Zharikov et al. ]{
G. Subebekova,$^{1}$
S. Zharikov$^{2}$\thanks{E-mail:zhar@astro.unam.mx} ,
G. Tovmassian$^{2}$,
V. Neustroev$^{3}$,
M. Wolf~$^{4}$,
\newauthor 
M.-S. Hernandez$^5$,
H. Ku\v{c}\'akov\'a~$^{4,6,7}$,  S. Khokhlov$^{1}$.
\\
$^{1}$Al-Farabi Kazakh National University, Al-Farabi Ave., 71, 050040, Almaty, Kazakhstan\\
$^{2}$Instituto de Astronom{\'i}a, Universidad Nacional Aut{\'o}noma de M{\'e}xico,
         Apdo. Postal 877, Ensenada, 22800 Baja California, M{\'e}xico\\
$^{3}$Space Physics and Astronomy research unit, PO Box 3000, FIN-90014 University of Oulu, Finland \\
$^{4}$Astronomical Institute, Faculty of Mathematics and Physics, Charles University, V~Hole\v{s}ovi\v{c}k\'ach~2, CZ-180~00~Praha~8,  Czech Republic \\
$^5$ Instituto de F\'isica y Astronom\'ia, Universidad de Valpara\'iso, Av. Gran Breta\~{n}a, 1111, Valpara\'iso, Chile \\
$^6$ Astronomical Institute, Academy of Sciences, 
	 Fri\v{c}ova 298, CZ-251~65, Czech Republic \\
$^7$ Research Centre for Theoretical Physics and Astrophysics, Institute of Physics, 
Silesian University in Opava, Bezru\v{c}ovo n\'am. 13, \\ CZ-746 01 Opava, Czech Republic
}

\date{Accepted XXX. Received YYY; in original form ZZZ}

\pubyear{2015}

\begin{document}
\label{firstpage}
\pagerange{\pageref{firstpage}--\pageref{lastpage}}
\maketitle

\begin{abstract}
We obtained photometric observations of the nova-like cataclysmic variable RW Tri 
and gathered all available AAVSO and other data from the literature. We determined the
system parameters and found their uncertainties using the code developed by us to model
the light curves of binary systems. New time-resolved optical spectroscopic observations of
RW Tri were also obtained to study the properties of emission features produced by the system. The
usual interpretation of the single-peaked emission lines in nova-like systems is related to
the bi-conical wind from the accretion disc's inner part. However, we found that the H$\alpha$
emission profile is comprised of two components with different widths. We argue that the
narrow component originates from the irradiated surface of the secondary, while the broader
component's source is an extended, low-velocity region in the outskirts of the accretion
disc, located opposite to the collision point of the accretion stream and the disc. It appears
to be a common feature for long-period nova-like systems -- a point we discuss.

\end{abstract}

\begin{keywords}
binaries: close; binaries: spectroscopic; stars: individual: RW Tri;  novae; cataclysmic variables; 
\end{keywords}



\section{Introduction}
\label{intro}

\begin{figure*}
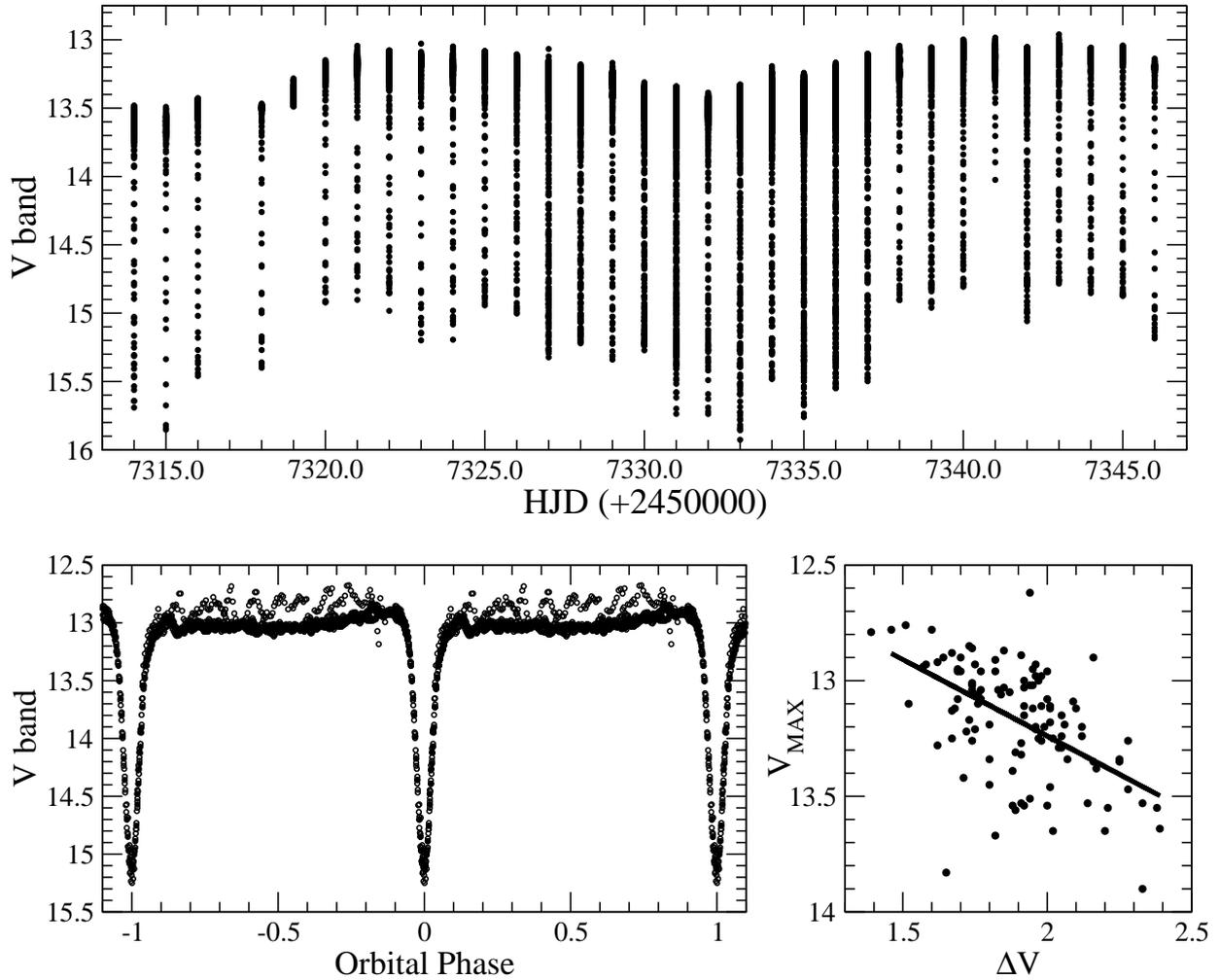

\setlength{\unitlength}{1mm}
\resizebox{15.cm}{!}{
\begin{picture}(130,117)(0,0)
\put(-10,0) {\includegraphics[width=14.5cm,angle=0,bb=20 30 780 390,clip=]{zharFig1a.eps}}
\put(-10,54.5) {\includegraphics[width=14.25cm,angle=0,bb=20 30 770 375,clip=]{zharFig1b.eps}}
\end{picture}}
    \caption{Top panel: Long-term light curve of RW Tri collected by AAVSO. Bottom-left: The example of the V band light curve 
    of  RW Tri folded on the orbital period. The data were obtained in four consecutive nights in OAN-SPM.
    The light curve with flickering is shown by the open circles and the opposite is presented by the filled circles.   
    The difference in out-eclipse behaviors for diverse orbital cycles is clearly visible. Bottom-right: The relationship between
    out-eclipse maximal brightness and the depth of the eclipse. }
    \label{lightAAVSO}
\end{figure*}

Cataclysmic variables (CVs) are interacting binaries comprised of a white dwarf (WD) as the primary and a late-type (K-M type)
main-sequence star or a brown dwarf as the secondary \citep{1995Ap&SS.226..187W}. The secondary star  fills its Roche
lobe and  looses matter via the inner Lagrangian point L$_1$ forming an accretion disc around the white dwarf, if the magnetic field of the white dwarf is not strong enough to prevent this. 
CVs which show outbursting behaviour related to an explosive transition in the disc between low and high temperature/density states are called dwarf novae (DNe). In contrast to them, there are systems at the long-period end of the orbital period distribution of CVs (P$_{\mathrm{orb}}>3$h), which exhibit a high accretion rate $\ge10^{-9}$M$_{\sun}$. Such systems are called nova-likes (NLs); their accretion discs stay in a high state most of the time. 
 Low-resolution optical spectra of NLs show a  blue continuum with single-peaked  Balmer emission lines, even for high inclination eclipsing systems. In some cases the emission lines are  flanked  by absorption features. 
The UV spectrum of NLs is dominated by emission from the accretion disc. They show resonance lines such as \ion{N}{V}, \ion{Si}{IV}, and \ion{C}{IV} often having  P~Cygni-like and/or blueshifted absorption profiles  \citep[and references therein]{2010ApJ...719.1932N}. Several  models (Stark broadening --
\citealt{1988ApJ...327..234L}; magnetic accretion -- \citealt{1989AJ.....97.1752W};  disc-overflow
accretion -- \citealt{1994ApJ...431L.107H}; an extended bright spot as the dominant source of emission lines --
\citealt{1997MNRAS.291..694D, 2014AJ....147...68T}; and wind emission --
\citealt{1986ApJ...302..388H, 1996Natur.382..789M, 2010ApJ...719.1932N,2011ApJ...736...17P,  2015MNRAS.450.3331M} have been suggested to explain NLs characteristics.
The wind model which is more developed than others is able to reproduce hydrogen and and helium recombination lines in the UV spectra of NLs, as well as a recombination continuum blue-wards of the Balmer edge.
But, the model  produces
stronger-than-observed \ion{He}{II} lines in the optical region and too much of a collisionally exciting contribution to the UV resonance lines.
It was reported that single-peaked profiles of Blamer lines in NLs probably have  complex structure \citep{1982ApJ...252..681H, 1983ApJ...267..239K,1994ApJ...424..347M, 2004MNRAS.353.1135T}.
Recent high-resolution spectral study of two NL systems  \citep{2017MNRAS.470.1960H}  demonstrates 
that Balmer emission line profiles  can be  divided into two components  arising from two distinct regions. The source of the narrow, low-velocity component is the irradiated face of the secondary
star. The wide component, according to  \citet{2017MNRAS.470.1960H}, is emanated from an outflow region in the vicinity of the L$_3$ point.

To verify this idea, we obtained new, time-resolved, high-resolution optical spectroscopic
observations of RW~Tri --- a well studied NL system.
Results of our spectroscopy combined with the new photometric data  
allowed us to redefine the system parameters of RW~Tri by using light curve modeling tool \citep{2013A&A...549A..77Z} and obtain additional evidence in favor of this hypothesis.

This paper is structured as follows. In Section~\ref{rwtri}, we describe previous progress in the study of RW~Tri.
Our spectroscopic observations of RW~Tri, 
the photometric data, and the corresponding data reduction are given in  Section~\ref{sec:obs}.  The photometric monitoring of RW~Tri, 
the optical light curve modeling, and the system parameters estimation is presented in   Section~\ref{sec:photMod}.
In Section~\ref{sec:Spec}, we analyze RW~Tri spectra and behaviours of its emission-lines.  The general discussion 
of the results and their application to NLs placed in a similar orbital period range are given in Section~\ref{sec:discus}. 
Our conclusions are presented in Section~\ref{sec:Concl}.

\section{RW Trianguli}
\label{rwtri}

RW Tri was discovered as an eclipsing binary system by \cite{1938AN....266...95P}. {\it  GAIA} distance of the system is 
315.5$\pm$5.0 pc \citep{2018A&A...616A...9L}.  It is in a good agreement with  previous HST parallax measurements $341^{-31}_{+38}$ pc \citep{1999ApJ...520L..59M}.
The first extensive photoelectric photometry in the $UBV$ system was reported by \citet{1963ApJ...137..485W}. He showed that the object besides eclipses varies in brightness at a long and short time-scale. 
The depth of eclipse depends on the brightness of the system, which itself does not present any periodicity either 
on long or short scales. The orbital period of the system is P$_{\mathrm{orb}}= 0.23188324\pm4\times10^{-8}$ days \citep{1978PASP...90..568A}. 
The light curve of RW Tri has a close similarity to UX UMa, a prototype of NLs.
\citet{1981MNRAS.195..825L} observed RW\,Tri at the near-infrared (nIR) $JHK$ bands. The nIR light curves 
showed a deep primary minimum whose depth decreases towards longer wavelengths, and a shallow secondary minimum, 
not seen at the optical wavelengths. Light curves allowed to estimate basic system parameters 
 listed in Table~\ref{tab:system}. 
\citet{1981MNRAS.195..227F} used infrared and visual light curves of  RW\,Tri during quiescence to fit simultaneously 
by a model in which a red dwarf eclipses an optically thick steady-state accretion disc. They obtained self-consistent 
system parameters for their model, also included in Table~\ref{tab:system}. 
\citet{1983ApJ...267..239K} conducted time-resolved
spectrophotometry of RW Tri to study the rapid evolution of the emission line spectrum as the accretion
disc undergoes occultation by the secondary star. The object was in a dim state at $B\approx13.3$ mag. \citet{1983ApJ...267..239K} found that profiles of Balmer lines show a double component
structure. The line profiles presented in the form of trailed spectra (see Fig.~8 to 10 by
\citet{1983ApJ...267..239K}) leave an impression of the presence of two anti-phased components that cross
at phase $\varphi=$0.5 and $\varphi=$0.0. 
\citet{1985MNRAS.216..933H} reported simultaneous high-speed $UBR$-photometry of RW Tri and the study 
of its accretion disc based on eclipse mapping. 
They propose that the disc is optically thick, with temperature ranging from 10\,000\,K at 
the outer edge to 40\,000\,K near the center. The radial temperature profile is consistent with the
expected $\mathrm{T}\varpropto \mathrm{R}^{-3/4}$ law for steady-state accretion. 

 \citet{1991AJ....102.1176R} revised the eclipse ephemeris 
 and estimated 
 the limit of the period rate change $\dot{\mathrm{P}}<5.6\times10^{-12}$. They also re-evaluated an upper limit of the mass 
 transfer rate to $\dot{\mathrm{M}} < 5.6\times 10^{-9}$ M$_{\sun}$ yr$^{-1}$.
 \citet{1992A&A...253..139R} presented phase-resolved spectropolarimetric observations of RW Tri and did not find any intrinsic
 linear polarisation to the binary system. They also concluded that a significant fraction of line emission is radiated from a source
 other than the accretion disc.

The first Doppler tomography in H$\beta$, \ion{He}{I} $\lambda$4471\,\AA\ and \ion{He}{II} $\lambda$4686\,\AA\ of RW Tri were  
presented by \citet{1994ApJS...93..519K}. 
\begin{table*}
   \caption{Estimations  of RW\,Tri system parameters.}
\centering
    \begin{tabular}{c|c|c|c|c|c|c|c|l}
    \hline
Reference                         & Distance   & $i$      & M$_1$      & M$_2$      & R$_2$      & $SpT_2$ & R$_{disk}^{out}$   & \\  
                               & pc         & degree   & M$_{\sun}$  & M$_{\sun}$  & R$_{\sun}$  &         &  R$_{\sun}$ & \\ \hline
 \citet{1981MNRAS.195..825L}   & 400        & $>80$    & 1.4-2.0    & 0.7        & 0.7-0.8    & K5V     &  0.7-0.8   & \\ 
 \citet{1981MNRAS.195..227F}   & 180(70)    & 82       & 1.3(0.7)   &0.47        & $0.56_{-0.02}^{+0.06}$& M0V & 0.76-0.94 & \\ 
  \citet{1983ApJ...267..239K}  &     -      & 67-75    & 0.64(1)    &0.76(2)     &    -       &  K5-K7 &     0.34(12)$^{o}$  &  \\
          "                    &     -      & 67-75    & 0.54(9)    &0.69(2)     &    -       &  K5-K7 &     0.22(5)$^{o}$  &  \\
 \citet{1992AA...260..213R}   &    -       &  75      &0.7        &  0.6       &      -      &     -   &      0.25$^{o}$   &  \\  
 \citet{1995AcA....45..259S}   & 240(40)    & 70.5     & 0.45(15)   &0.63(15)    &     -       &    -    &     0.25  &  \\    
 \citet{2003MNRAS.340..499P}   &    -       &  -       &0.4-0.7     & 0.3-0.4    &     -       &    -    &      -     &  \\   
 \citet{2019AcA....69...79S}   &    -       &  72.5(2.5)      &0.6(2)     & 0.48(15)    &    -        &    -    &    -       &  \\  \hline 
 
\end{tabular}
\begin{tabular}{l}
 $^{o}$ - optical radius of the accretion disc \\
 $SpT_2$ - spectral type of the secondary
 \end{tabular}
    \label{tab:system}
\end{table*} 
\begin{figure*}
\setlength{\unitlength}{1mm}
\resizebox{15.cm}{!}{
\begin{picture}(130,117)(0,0)
\put(-25,-5) {\includegraphics[width=14.7cm, bb= 0 120 650 650, clip=]{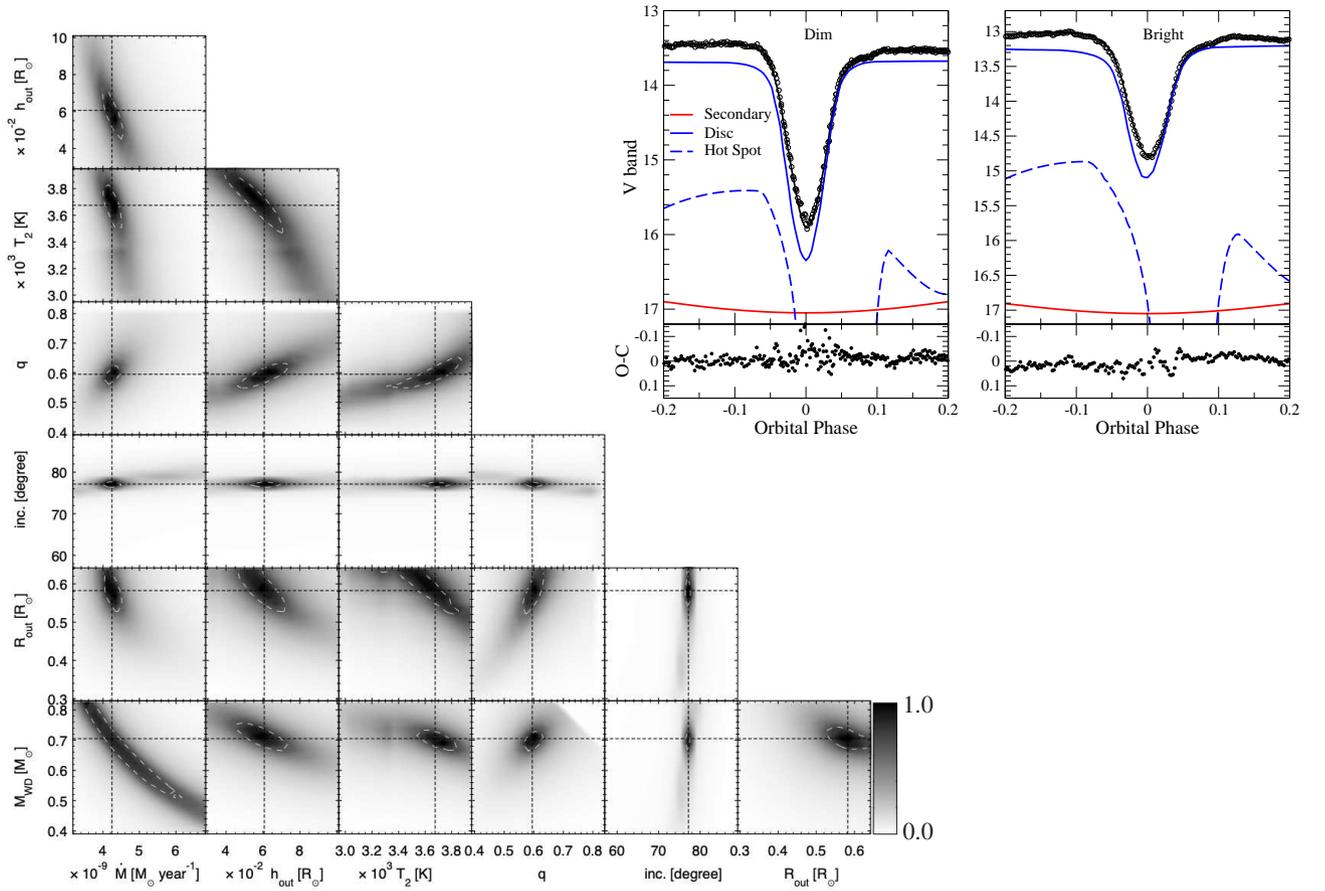}}
\put(62,55) {\includegraphics[width=8.cm,angle=0, bb=30 50 780 565,clip]{zharFig2a.eps}}
\put(96,28){\small 1.0}
\put(96,13.5){\small 0.0}
\end{picture}}
    \caption{Right-top panels: Examples  of eclipses (open circles)  in the `bright' (right) and in the `dim' (left) state of 
    the system and results of the light curve modeling (solid line). Contribution  of different system components in the
    total flux are marked. The corresponding  $O-C$ (observed minus calculated) diagrams are given in the bottom panels of this plot.
    Left gray scale panels: Errors of the fit for the light curve in the `dim' state. The black dashed lines mark 
    the best fit values of parameters.  The white long-dashed lines corresponds 1$\sigma$ errors of parameters. }
    \label{fig:LCmodel}
\end{figure*}
\citet{1995MNRAS.273..849S} carried out time-resolved spectrophotometry 
of RW Tri intending to define the regions of optical emission. They found that the variation of the Balmer line profile can be
interpreted as a two-component emission. 
There is no evidence of an eclipse in the emission wings of these lines at phase zero, and there is no suggestion of a rotational
disturbance in either the broad emission component or the absorption wings, expected from a rotating accretion disc.

RW Tri was observed with the Goddard High-Resolution Spectrograph (GHRS) with the {\it HST}
in a program primarily designed to study accretion disc winds \citep{1997MNRAS.290L..23M}. The light curve of the object exhibits a
prominent UV eclipse. It is U-shaped and 
relatively wide (full width at half-light, $\Delta\varphi_{0.5} = 0.07$ cycles). The
most interesting feature of the reported  UV light curve is a dip in the UV flux during the half-cycle preceding the eclipse. 
The dip shows the maximum depth at the phase $\varphi\approx0.7$. It has a distinct saw-tooth shape with a rapid fall beginning at  phase of $\varphi\sim0.5-0.6$
and a more gradual rise, which is not complete until the phase of $\varphi\sim0.2$ (i.e. post-eclipse).

The secondary in RW\,Tri is probably M-type main-sequence  star judging  from the $I$- and $K$- band  spectroscopy \citep{2000MNRAS.314..826D,
2003MNRAS.340..499P}. 
The $K$-band light curve reaches $K=12.17$ at the primary minimum
\citep{1981MNRAS.195..825L}, which combined with the distance results in $M_{K_s}\geq4.65$. It limits  the spectral class of the
secondary to later than K6V or $\mathrm{T}_2 \leq 4250$\,K. However, much deeper eclipse 
to  $V=16.23\pm0.07$  establishes the upper limit to $M_V\geq8.47$ or T$_2 \leq 4000$\,K
\citep{2013ApJS..208....9P}.
\citet{2003MNRAS.340..499P}  made an effort to determine the orbital velocity of the secondary using time-resolved spectroscopy in
$I$ and $K$- bands by skew-mapping and cross-correlation techniques. They found a radial velocity amplitude of  $250\pm47$
km~s$^{-1}$ in the $I$- band, and $229\pm29$km~s$^{-1}$ in the $K$- band. 
A combination of these results leads to the best estimations
of the primary and the secondary stellar masses in the range of 0.4-0.7 and 0.3-0.4 M$_{\sun}$, respectively. 
The last determination
of the system parameters was reported by \citet{2019AcA....69...79S} based on new spectroscopic results and old photometric data
where negative superhumps were detected.

In Table~\ref{tab:system} we summarise the results of system parameters determination in previous studies of RW Tri.
As can be seen, despite a large amount of data and many carried out studies, there is a significant spread of the determined parameter values. This encouraged us to attempt to refine them using the light curve modeling technique developed by us (see below).

\section{OBSERVATIONS AND DATA REDUCTION}
\label{sec:obs}
A time-resolved CCD photometry of RW Tri 
has been obtained at the Ond\v{r}ejov Observatory,
Czech Republic, since 2012.  The Mayer 0.65 m ($f/3.6$) reflecting telescope with the CCD camera G2-3200 and \textit{VR} photometric filters
were used. {\sc Aphot}, synthetic aperture photometry, 
and astrometry software were used for data reduction. Differential photometry was performed using suitable comparison 
stars, and a heliocentric correction was applied. Computers at the telescope are synchronized using a time-server provided
by {\tt tick.usno.navy.mil} every two minutes. These corrections are usually of the order of $10^{-3}$ seconds.

In addition,  time-resolved $V$-band CCD photometric data were obtained 
on 2016 September 8-11,  and on 2016 November 14, 16, and 17 using 0.84m telescope and MEXMAN
filter-wheel of Observatorio Astronomico Nacional at San Pedro M\'{a}rtir (OAN-SPM), Mexico.

The spectroscopic data of RW Tri were obtained using the echelle REOSC spectrograph \citep{1995Levin} and 
 Boller \& Chivens long-slit spectrograph attached to the 2.1-m
telescope of the OAN-SPM. The echelle spectrograph provides spectra spread over 27 orders, covering the spectral range 3500-7100 \AA\ with 
the spectral resolving power of R=18000.  A total of 79 echelle spectra were obtained in 2016, September, and October. 
The echelle spectroscopy was obtained simultaneously with time-resolved 
photometric observations. 

Also, a number of low-resolution spectra were obtained using Boller \& Chivens long-slit spectrograph.
The Boller \& Chivens covers spectral range 3900-7400 \AA\ with 
the spectral resolving power of R$\approx$2000. All spectroscopic observations were obtained under photometric conditions. 
Standard procedures, including bias and flat-field correction, cosmic ray removal, and wavelength calibration, were applied.
The log of spectroscopic observations is presented in Table~\ref{tab:LogSpObs}.

\begin{table}
\label{tab:LogSpObs}
\caption{Log of spectral observations of RW\,Tri.}
    \begin{tabular}{c|l|c|c|c}
    \hline
Date   &  HJD start  & Number & Exp. Time & Duration \\
DD/MM/YYYY & +2450000& of exp. & (s)      &(h)      \\
   \hline
08/09/2016 & 7639.8046 &  15  & 1200     &  4.9    \\
09/09/2016 & 7640.7895 &  16  & 1200     &  5.2    \\
10/09/2016 & 7641.7732 &  17  & 1200     &  5.7    \\
11/09/2016 & 7642.7726 &  17  & 1200     &  5.6    \\
12/09/2016 & 7643.7624 &  14  & 1200     &  4.6    \\
27/09/2016 & 7658.9281$^*$ &  6   & 240      &  0.4       \\
14/10/2017 & 8040.9466$^*$ &  6   & 900      &  1.5       \\
   \hline
\end{tabular}
\begin{tabular}{l}
    Spectra were obtained with the OAN SPM REOSC echelle spectrograph \\ except for those marked by
    $^*$ for which B\&Ch spectrograph was used.
\end{tabular}
\end{table}

\section{PHOTOMETRY AND THE LIGHT CURVE MODELLING OF RW TRI}
\label{sec:photMod}

\begin{table}
\caption{System parameters used in the light curve modelling.}
\label{tab:BestPar}
\begin{tabular}{llll}
\hline \hline
{\bf Fixed:}       &                                  &              &         \\ \hline
P$_{\mathrm{orb}}$          &     20034.72s                    & $E(B-V)$     & 0.10    \\
Distance           &     315(5) pc                    &  $V_2 \sin (i) $           &    221(29) km s$^{-1}$       \\ \hline \hline

\multicolumn{4}{l}{{\bf Variable and their best values:}} \\ \hline
\multicolumn{2}{l}{ $i$ }              & \multicolumn{2}{c}{ 77.2(5) } \\
\multicolumn{2}{l}{M$_{\mathrm{WD}}  $ }  &\multicolumn{2}{c}{0.70(3) M$_{\sun}$}  \\
\multicolumn{2}{l}{R$_{1}$ }     & \multicolumn{2}{c}{0.011  R$_{\sun}$}\\
\multicolumn{2}{l}{$q $}         & \multicolumn{2}{c}{$0.60(3)$} \\
\multicolumn{2}{l}{M$_{2}  $}    & \multicolumn{2}{c}{ 0.42 M$_{\sun} $} \\
\multicolumn{2}{l}{$\mathrm{T}_{2}$ }     & \multicolumn{2}{c}{3675(125)  K}\\
\multicolumn{2}{l}{R$_{2}$ }     & \multicolumn{2}{c}{0.55  R$_{\sun}$}\\
\multicolumn{2}{l}{ R$_{\mathrm{disc}}^{\mathrm{out}}$ } & \multicolumn{2}{c}{ 0.58(4) R$_{\sun}$ } \\ \hline
\multicolumn{2}{l}{\bf dim}    & \multicolumn{2}{c} {} \\
\multicolumn{2}{l}{$\dot{\mathrm{M}}$} & \multicolumn{2}{c}{4.3(2)$\times10^{-9}$M$_{\sun}$ year$^{-1}$}\\
\multicolumn{2}{l}{ h$^{\mathrm{out}}_{\mathrm{disc}}$ } & \multicolumn{2}{c}{ 0.06(1)  R$_{\sun}$ } \\ \hline
\multicolumn{2}{l}{\bf bright}    & \multicolumn{2}{c} {} \\
\multicolumn{2}{l}{$\dot{\mathrm{M}}$}    & \multicolumn{2}{c}{7.8$\times10^{-9}$M$_{\sun}$ year$^{-1}$}\\
\multicolumn{2}{l}{ h$^{\mathrm{out}}_{\mathrm{disc}}$ } & \multicolumn{2}{c}{0.10  R$_{\sun}$ }\\ \hline
\multicolumn{2}{l}{$a  $ }  &\multicolumn{2}{c}{1.65  R$_{\sun}$}  \\
\multicolumn{2}{l}{$V_1 \sin(i)  $ }  &\multicolumn{2}{c}{135 km  $s^{-1}$}  \\ 

 \hline 

\end{tabular}
\begin{tabular}{l}
Numbers in brackets throughout the paper  are $1\sigma$ 
uncertainties \\ referring to the last significant digits quoted.
\end{tabular}
\end{table}

\begin{table}
\begin{center}
\caption{Precise times of RW~Tri eclipses measured  in Ondrejov during 2012-2020.}
\label{tmin}
\begin{tabular}{llcl}
\hline\hline
 HJD -    &  Error  & Filter & Epoch      \\
 2400000  &  [day]  &        &         \\ 
\hline
  56527.56969  &  0.0001  &  $R$  &  66405  \\ 
  56540.55508  &  0.0001  &  $R$  &  66461  \\ 
  56624.26501  &  0.0001  &  $R$  &  66822  \\ 
  56659.27939  &  0.0001  &  $R$  &  66973  \\  
  56870.52515  &  0.0001  &  $R$  &  67884  \\ 
  56926.40887  &  0.0005  &  $R$  &  68125  \\ 
  57059.27813  &  0.0001  &  $R$  &  68698  \\   
  57295.33559  &  0.0001  &  $R$  &  69716  \\ 
  57327.33545  &  0.0001  &  $R$  &  69854  \\ 
  57384.37895  &  0.0001  &  $R$  &  70100  \\ 
  57720.37804  &  0.0001  &  $R$  &  71549  \\ 
  57722.23322  &  0.0001  &  $V$  &  71557  \\ 
  57739.39218  &  0.0001  &  $V$  &  71631  \\ 
  57780.20443  &  0.0001  &  $R$  &  71807  \\ 
  57798.29077  &  0.0001  &  $R$  &  71885  \\ 
  57958.52201  &  0.0001  &  $R$  &  72576  \\ 
  57994.46451  &  0.0001  &  $V$  &  72731  \\ 
  58004.43543  &  0.0001  &  $V$  &  72774  \\ 
  58041.53636  &  0.0001  &  $V$  &  72934  \\   
  58089.30455  &  0.0001  &  $V$  &  73140  \\  
  58149.36246  &  0.0001  &  $V$  &  73399  \\ 
  58169.30443  &  0.0001  &  $V$  &  73485  \\ 
  58185.30428  &  0.0002  &  $V$  &  73554  \\  
  58326.52106  &  0.0002  &  $V$  &  74163  \\  
  58329.53560  &  0.0002  &  $V$  &  74176  \\ 
  58366.40488  &  0.0002  &  $V$  &  74335  \\ 
  58387.50619  &  0.0002  &  $V$  &  74426  \\ 
  58429.24511  &  0.0001  &  $V$  &  74606  \\ 
  58458.23023  &  0.0001  &  $V$  &  74731  \\  
  58502.28782  &  0.0001  &  $V$  &  74921  \\ 
  58505.30230  &  0.0001  &  $V$  &  74934  \\ 
  58506.22975  &  0.0001  &  $V$  &  74938  \\ 
  58527.33128  &  0.0001  &  $V$  &  75029  \\ 
  58710.51838  &  0.0001  &  $V$  &  75819  \\ 
  58714.46076  &  0.0002  &  $V$  &  75836  \\ 
  58773.59019  &  0.0001  &  $R$  &  76091  \\ 
  58773.59025  &  0.0001  &  $V$  &  76091  \\ 
  58828.31536  &  0.0001  &  $R$  &  76327  \\ 
  58828.31551  &  0.0001  &  $V$  &  76327  \\ 
  58866.34418  &  0.0001  &  $R$  &  76491  \\ 
  58866.34411  &  0.0001  &  $V$  &  76491  \\ 
  58892.31521  &  0.0001  &  $R$  &  76603  \\ 
  58892.31517  &  0.0001  &  $V$  &  76603  \\ 
  \hline
\end{tabular}
\end{center}
\end{table}

A fragment of about a one-month-long $V$-band light curve of RW Tri based on the AAVSO data is plotted in Figure~\ref{lightAAVSO}.
Besides eclipses, the object shows long-term variability in a 12.8 -- 13.7 range of magnitudes. A zoom on one orbital period is
presented at the bottom-left panel, on two different occasions (overplotted), when the flickering is present (open circles) and 
when it is absent (filled circles; the light curve per period is repeated twice). The flickering appears and disappears randomly. 
The eclipse depth ($\Delta V$) depends on the underlying
brightness ($V$) of the object  (see the bottom-right panel of Figure~\ref{lightAAVSO}).  All these features are generally well known (see Section~\ref{intro}), but the deduced system parameters demonstrate a large spread (see Table~\ref{tab:system}).

We used gathered photometric observations to analyze eclipse light curves 
using a tool developed by \citet{2013A&A...549A..77Z} to  improve the system's definition. Briefly, the model  includes a primary white dwarf, a secondary red dwarf star, a stream of accretion matter, thick $h(r) = h(r_{out})(r/r_{out})^\gamma$ ($\gamma=1$) accretion disc, and an extended hot spot/line. The white
dwarf is a sphere, defined by the mass-radii relation in \cite[2.83b]{1995Ap&SS.226..187W}. The secondary is assumed to fill its Roche lobe,
and the Roche lobe shape is directly calculated using Equation
\cite[2.2]{1995Ap&SS.226..187W} for equipotential $\Phi$(L$_1$). The surface of each component of the system is divided into a series of triangles. We assume that each triangle emits as a blackbody with the corresponding temperature. The limb-darkening \citep{2012A&A...546A..14C} and 
the illumination of the secondary by the primary are also included. Each element's intensity is convolved with the corresponding filter bandpass and converted into the flux taking into account the element surface, orientation,  distance to the system, and interstellar absorption.   The light curves of individual components and  the binary system as a whole were obtained
by integrating the emission from all the elements lying in view. 
Details of the light curve modeling of similar NLs are described in \citet{2014AJ....147...68T, 2017MNRAS.470.1960H}.

\begin{figure}
\setlength{\unitlength}{1mm}
\resizebox{15.cm}{!}{
\begin{picture}(130,50)(0,0)
\put(0,0) {\includegraphics[width=7.5cm, bb=10 20 790 530, clip=]{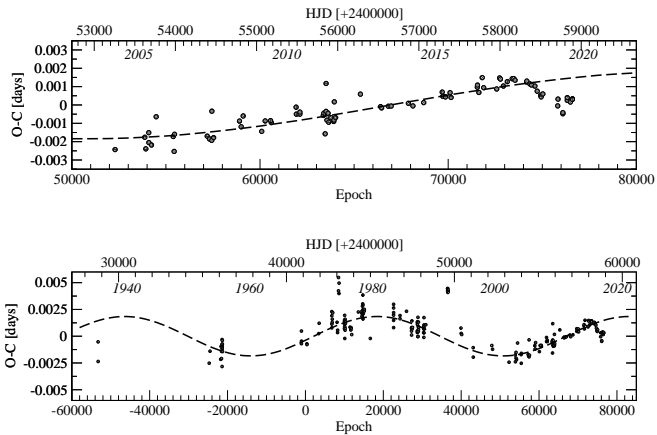}}
\end{picture}}
\caption[ ]{Bottom panel represents the historical \oc\ diagram for the mid-eclipse times of RW~Tri. 
Individual minima are denoted by circles. The dashed sinusoidal curve represents a possible period  (42.2 years) present in \oc\ variation.
At the top the \oc\ diagram of the ultimate decade is presented  in detail based
mostly on our new measurements from Ond\v{r}ejov. A rapid decrease of the orbital period since 2018 is visible. }
\label{rwoc}
\end{figure}

 After revising a large volume of the AAVSO data, and our proper observations, we selected examples of two different  eclipses of RW~Tri presented in the upper-right corner of Figure~\ref{fig:LCmodel}. However, we note that these two extreme examples represent typical eclipse profiles in a bright and dim state, and many equivalents can be found at different epochs in the  AAVSO data. In the right panel of Figure~\ref{fig:LCmodel}, the eclipse profile with the smallest eclipse depth is presented 
(which occurs when the system 
is brighter). The plot in the left panel corresponds to one of the deepest observed eclipses during the overall decrease of the
object brightness.   They both were fitted with congruent models. 
The fit aims to determine system parameters of RW~Tri, given that it 
is a high inclination system located at a distance of 315\,pc. We also have the measure of the secondary orbital velocity
K$_2$ = V $\sin$(i) =  200-240 km~s$^{-1}$.  In addition, we assume that a standard steady-state accretion  defines the disc temperature radial profile $\mathrm{T}\propto \mathrm{R}^{-3/4}$ with an accretion rate of  $\dot{\mathrm{M}} \approx 10^{-9}-10^{-8}$ M$_{\sun}$~yr$^{-1}$. 
The last assumption is based on the previous study of the system (see Section~\ref{intro}), our  low-resolution spectroscopy, 
and the object brightness. 

Free parameters of the fit are the mass of the primary (M$_{\mathrm{WD}}$), the mass ratio ($q\equiv \mathrm{M_2/M_{WD}}$), the mass transfer rate
($\dot{\mathrm{M}}$),  the system inclination ($i$), the outer radius of the accretion disc (R$_{\mathrm{disc}}^{\mathrm{out}}$),
the disc height at the outer radius (h$_{\mathrm{disc}}^{\mathrm{out}}$), and the effective temperature of the secondary ($\mathrm{T}_2$).
The formal temperature of the primary white dwarf was selected to be 30\,000\,K. It can be significantly higher because the
dominating source of radiation is the inner part of the accretion disc. The variation of the white dwarf temperature in the range of 18\,000
- 35\,000K  does not affect the eclipse profile. We fixed parameters of the hot spot (length, width, and temperature) for the bright and the dim light curves to reproduce a small peak observed at the
orbital phase of $\varphi\approx 0.12$.

 The gradient descent method was used to find the minimum of the $\chi^2$ function defined as 
\begin{equation}
 \chi^2 = \sum_k^{N_k}\left(\frac{mag^{obs}_k-mag^{calc}_k}{\sigma~mag^{obs}}\right)^2     
\end{equation}

$N_k$ is the number of observed points in each fitted light curve.

The values of the best fit model are presented in Table~\ref{tab:BestPar}. A mosaic of square-shaped panels on the left side of
Figure~\ref{fig:LCmodel}  illustrates errors of the fit attained for the deep eclipse in the dim state of brightness. The light
curve in a brighter state  was reproduced with the same system parameters, but with a higher value of mass transfer rate than in 
the dim state. The outer radius of the disc R$_{\mathrm {disc}}^{\mathrm {out}}=0.58$R$_{\sun}$ is close to the tidal limiting radius
$R^{\mathrm{max}}_{\mathrm {disc}}=\frac{0.6}{1+q}a=0.62$R$_{\sun}$. It is consistent with the  R$_{\mathrm {disc}}^{\mathrm{out}}=0.53-0.63$R$_{\sun}$ estimate from the nIR photometry of \citet{1981MNRAS.195..825L} recalculated for a new {\it GAIA} distance.
This size of the disc  provides the minimum velocity of the Keplerian disc of
$ \upsilon_{\mathrm {min}}\sin (i) = 470$ km s$^{-1}$. This velocity is shown by a circle in the Doppler maps in Figure~\ref{DopTom}.
The system inclination $i =77\fdg2\pm0.5$ is very well constrained  and is consistent with the   77\fdg5 value deduced by \citet{1981MNRAS.195..825L} from the eclipse width in the nIR. 
The temperature at the outer radius of the disc, according to our model, is about 5000\,K, and the maximum temperature of the hot spot is $\sim$7100\,K. The inner temperature in the disc reaches up to $\sim$35\,000\,K.  
The deduced secondary effective temperature T$_2=3675(125)$\,K corresponds to  M1.5V spectral type.
However, its radius is about 20\% larger than a corresponding ZAMS star. 
The main difference between the bright and the dim states in our model, apart from the change of the mass transfer rate, is related
to the vertical width of the accretion disc at the edge.  In general, all observed optical eclipse light curves can be reproduced with reasonable errors using the best-fit values with a variation of those two parameters.

We fitted the nIR  $J$- and  $K$-band light curves published by  \citet{1981MNRAS.195..825L} using 
our models for the dim and the bright states. 
We found that our model successfully reproduces observed nIR light curve shapes. However, the calculated flux is about 0.5 magnitudes fainter than it was reported by \citet{1981MNRAS.195..825L}. An excess in IR fluxes for wavelengths longer than $\sim$3-5 $\mu$m over expected from the standard model accretion disc was noted previously by \citet{2014ApJ...786...68H}  for the sample of 12 NLs. Therefore, we do no exclude IR excess that may appear  at a shorter wavelength range.   The size and shape of the eclipse are different in the UV, optical, and IR bands. If in UV domain, the full eclipse has a U-shape and lasts $ \approx 0.07 $P$_{\mathrm{orb}}$ \citep{1997MNRAS.290L..23M},  in  optical and nIR bands, it is V-shaped with a full time of $ \approx 0.14 $P$_{\mathrm{orb}}$, and $ \approx 0.2 $P$_{\mathrm{orb}}$  respectively \citep{1981MNRAS.195..825L}. It is clear that the disk is larger in the IR, not exactly circular or symmetric regarding the view of sight \citep[see for examples][ and references therein]{1982A&A...114..319H, 1993MNRAS.264..691M, 1996MNRAS.279..402M, 2001MNRAS.322..499K, 2017ApJ...841...29J}. Most probably, the outer fringe of the disc is ragged and not even in a vertical extend. All these departures from the ideal models might play a role in deviations of infrared flux, detection of a dip at the orbital phase of $\varphi\sim0.7$ in the UV light curve, and other inaccuracies compared to our and other models.

\begin{figure*}
\setlength{\unitlength}{1mm}
\resizebox{15.cm}{!}{
\begin{picture}(130,160)(0,0)
\put(-10,0) {\includegraphics[width=13.5cm,angle=0,bb = 50 10 770 880, clip=]{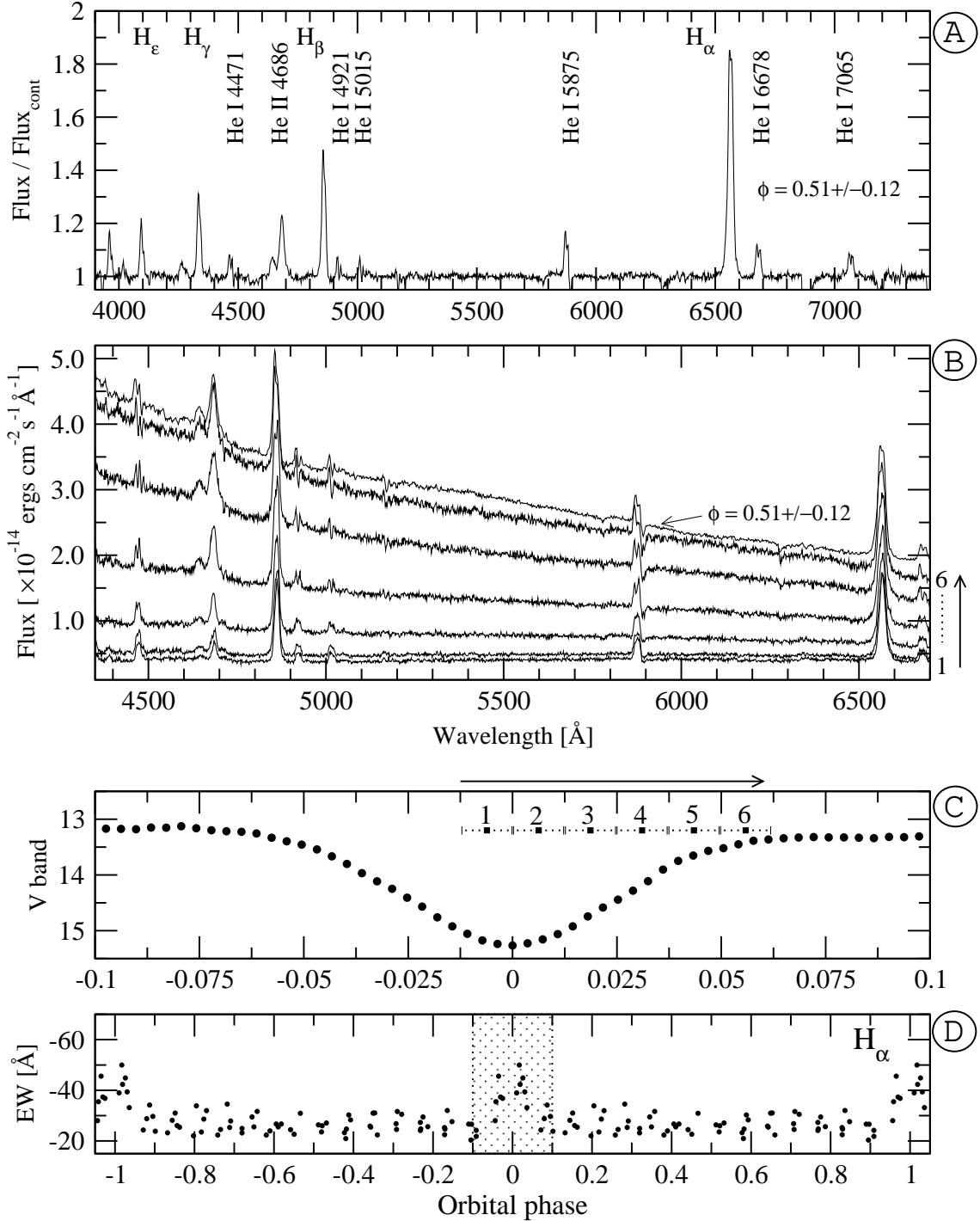}}
\end{picture}}
    \caption{A: Normalized low-resolution spectrum  of RW Tri averaged over a stretch of $\varphi = 0.47-0.73$ orbital phases.
    B: Spectral evolution of  RW Tri from the bottom  to clearing of the eclipse. The final (top) spectrum corresponds to the average spectrum for $\varphi =0.47-0.73$ orbital phases.
    C: The eclipsing light curve of RW Tri in the $V$-band obtained close to the spectral observations. The filled  squares mark the orbital phases when the spectra were acquired.
    D: The behaviour of equivalent widths (EW) of the H$\alpha$ emission line during the orbital period. The selected region corresponds to the panel C abscissa range. }
    \label{LowSpec}
\end{figure*}

\subsection{\oc\ diagram}

Since its discovery, RW~Tri has been continuously monitored, and many precise mid-eclipse epochs were determined. Thus its orbital period is known with a high precision of the order of $\sim10^{-3}$s 
\citep{1963ApJ...137..485W, 1978PASP...90..568A, 1991AJ....102.1176R}.  
Using high-time precision photometric data obtained in Ond\v{r}ejov Observatory, we calculated the time and uncertainties of each full eclipse by the least-squares Gaussian fitting of eclipse profiles.
The  predicted epochs of eclipses  were calculated using the ephemeris of \citep{1991AJ....102.1176R}:
 \begin{equation}
 \mathrm{T_{min}} = \mathrm{HJD} ~24 41129.36487(10) + 0\fd231883297(6) \cdot \mathrm{E}
 \end{equation}
\noindent
They are given in Table~\ref{tmin}.  The search for \oc\  periodic variation
 was performed using all available mid-eclipse times found in the literature (\oc\ gateway\footnote {\tt http://var2.astro.cz/ocgate, \citet{2006OEJV...23...13P}}), including our new results\footnote{Besides those minima included in Table~\ref{tmin}, we used the previous collection of eclipse times obtained by
\cite{1963ApJ...137..485W},
\cite{1973IBVS..852....1W},
\cite{1977AJ.....82.1008W},
\cite{1978PASP...90..568A}, 
\cite{1979AcA....29..469S},
\cite{1991AJ....102.1176R},
\cite{1992BCrAO..86...58D},
\cite{1992AA...260..213R},
\cite{2006IBVS.5710....1P},
\cite{2006IBVS.5741....1Z},
\cite{2012JAVSO..40..295B},  
and many other amateur observers.}. A total of 221 reliable times of minimum light of different weights were gathered.
The historical \oc\ diagram is plotted in Figure~\ref{rwoc}, bottom panel.
Obviously, mid-eclipse times do not follow a simple linear or parabolic trend. 
The data can be fitted roughly by a sinusoidal curve with a period of $\sim$41.2 years and the semi-amplitude of 160 sec. This result is consistent with the \oc\ behaviour recently reported by \citet{2012JAVSO..40..295B}.  If this periodicity is real, it might be related to the presence of a third body in the system at a wide orbit. For example, the third body in a similar system, LX~Ser, was recently reported by \citet{2017PASJ...69...28L}. An alternative explanation of cyclic period changes in RW~Tri could be the Applegate mechanism \citep{1992ApJ...385..621A} due to the magnetic activity of the red dwarf component. According to this mechanism, a solar-like magnetic cycle would result in shape changes causing redistribution of the angular momentum within the interior of the secondary. The change of the stellar quadrupole moment then leads to the variation of the orbital period. 
Our estimation of the \oc\ periodic variation with $P \approx 42.2$ years is comparable with the mean value (40-50 years) of quadrupole moment changes/modulation in magnetically active close binaries \citep{1999A&A...349..887L}.

However, the periodicity is not an established fact, and more observations are necessary to confirm its reality.
A portion of \oc\ diagram, based only on our recent precise-timing data, is plotted in the top panel of Figure~\ref{rwoc}. Since 2018, the orbital period abruptly decreases, deviating from the sinusoidal trend marked by the dashed line.

\begin{figure}
\setlength{\unitlength}{1mm}
\resizebox{15.cm}{!}{
\begin{picture}(130,60)(0,0)
\put(-2,0) {\includegraphics[width=7cm,angle=0,bb=70 45 651 565,clip]{zharFig5a.eps}}
\put(65,5)  {\includegraphics[width=1.cm,  angle=0,bb = 255 333 356 420, clip]{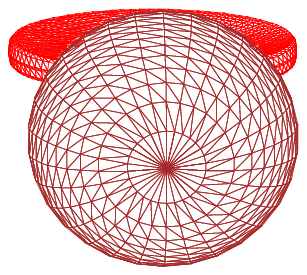}}
\put(64,28) {\includegraphics[width=1.22cm,angle=0,bb = 237 340 375 420, clip]{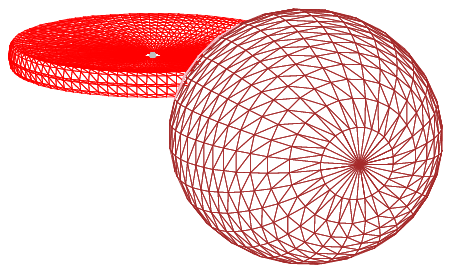}}
\put(65,45) {\includegraphics[width=0.95cm, angle=0,bb = 255 360 350 450, clip]{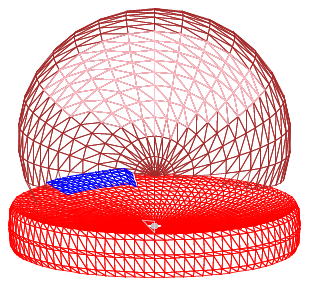}}
\end{picture}}
    \caption{Evolution of \ion{He}{I} line profiles during the eclipse (marked by 1 to 6) and at the orbital phase $\varphi = 0.5$.
    The continuum removed from the spectra after they were shifted along the y-axis
    for the best presentation. The right side of the plot presents an artistic view on the system at its corresponding orbital phases.}
    \label{fig:HeI}
\end{figure}

\begin{figure}
\setlength{\unitlength}{1mm}
\resizebox{15.cm}{!}{
\begin{picture}(130,60)(0,0)
\put(-3,0) {\includegraphics[width=9.3cm,angle=0,bb=0 30 870 780,clip]{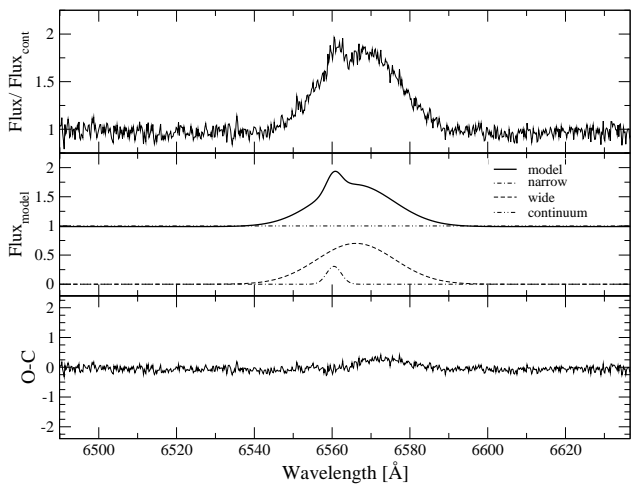}}
\end{picture}}
    \caption{An example of the high-resolution H$\alpha$ profile of RW~Tri (top panel), the result of a double-Gaussian fit to the profile (middle panel), and the residuals between the observed and calculated profiles (bottom panel). The spectrum was normalized to the continuum.}
    \label{fig:Ha}
\end{figure}

\begin{figure*}
\setlength{\unitlength}{1mm}
\resizebox{15.cm}{!}{
\begin{picture}(130,100)(0,0)
\put(-10,0) {\includegraphics[width=15cm,angle=0,bb = 0 390 600 1000, clip]{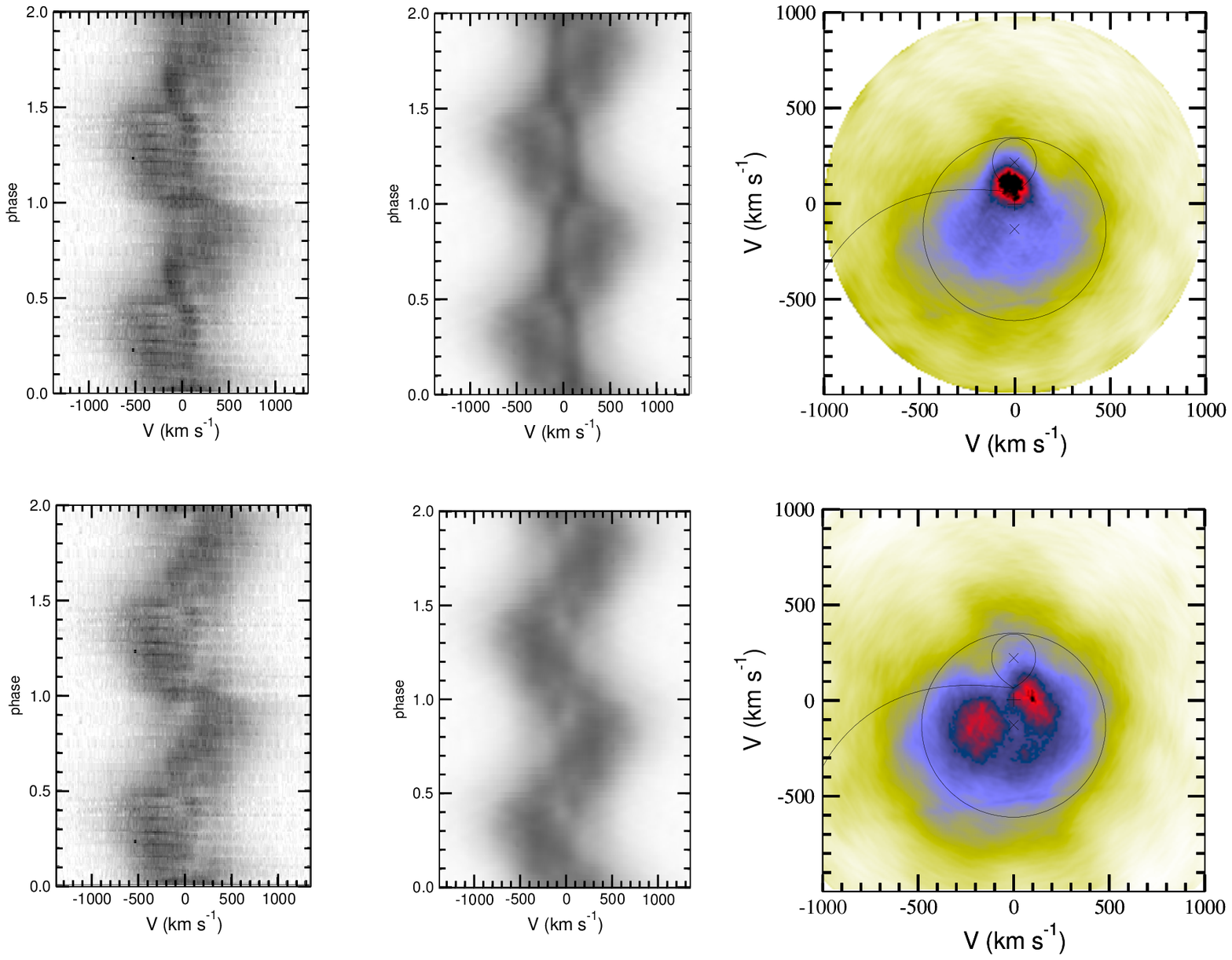}}
\end{picture}}
    \caption{Top panels from left to right:  The observed, reconstructed trailed spectra around H$\alpha$ emission line and corresponding Doppler tomogram. 
    Bottom: The observed, reconstructed, and Doppler tomogram after removing the narrow component of the emission line.
    The Keplerian velocity of the  disc in the Doppler maps is  located at $ \upsilon_{\mathrm {disc}}\sin (i) \ge 470$ km s$^{-1}$ and the circle shows the disc external radius. The system parameters discussed in the text.}
    \label{DopTom}
\end{figure*}

\section{Spectroscopy and Doppler tomography}
\label{sec:Spec}

\subsection{Low-resolution spectroscopy}

Low-resolution spectra of RW~Tri throughout eclipse and outside of it are shown in Figure~\ref{LowSpec}.
In those spectra, the Balmer lines have single-peaked emission profiles, but not regular Gaussian shapes. Meanwhile,  
multiple \ion{He}{I} lines in the observed range are double-peaked in contrast to the relatively strong but single-peaked \ion{He}{II} $\lambda$4686\AA\ line. There is no evidence of underlying absorption or P~Cyg-like features in Balmer lines frequently observed in other NL systems. The out-of-eclipse spectrum shows a blue continuum with a slope of $\alpha \approx -2.4$ where $F\propto\lambda^{-\alpha}$, close to the standard value of $-$2.33 for the steady-state disc (see Figure~\ref{LowSpec}, panels A and B). 
The 
ratio of equivalent widths of Balmer lines is
EW =   H$\alpha$:H$\beta$:H$\gamma$:H$\delta$:H$\varepsilon$ = 23.97:9.22:6.74:3.36:2.56 \AA. 
At the bottom of the eclipse, the spectrum of the source is rather flat, with its blue extreme gradually increasing towards eclipse end, as it is clearly seen from a sequence of six low-resolution spectra obtained on 2016 September 27  (see Figure~\ref{LowSpec}, panels B and C). The first two spectra correspond to the moments of minimum in the eclipse. All lines, including  \ion{He}{I} are single-peaked at a depth of eclipse with FWHM$\approx$800~km~s$^{-1}$. All emission lines are visible during the deepest eclipse phase. As the inner parts of the accretion disc open up, flux grows with the slope of the continuum becoming bluer,  and the intensities of emission lines start to increase, especially \ion{He}{II} (see Table~\ref{LineParEcl}). The shape of the \ion{He}{II} $\lambda$4686\AA\ line profiles looks single-peaked in all orbital phases. 

Meanwhile, \ion{He}{I} transforms into a double-peaked and the Balmer lines show complex structure. The EWs of H$\alpha$ and \ion{He}{I}  decrease significantly as the system leaves the eclipse in contrast to \ion{He}{II} $\lambda$4686\AA. The evolution of  EW$_\mathrm{{H\alpha}}$ during the full orbital period is shown at the bottom panel of Figure~\ref{LowSpec} (measured from the high-resolution Echelle spectra). It is clear that  EW$_{\mathrm{H\alpha}}$ stays constant in a range of values and only increases during the eclipse as most of the radiation from the accretion disc falls.

Especially noteworthy is the evolution of \ion{He}{I} throughout the eclipse (see  Figure~\ref{fig:HeI}). It is drastically different from other lines.  At the moment of the full eclipse, the line has a single-peak, somewhat boxy shape centered at about zero radial velocity. 
 As the accretion disc re-appears from behind the secondary, the line profile obtains two nearly symmetric sharp peaks.  In other words,  a strong absorption component appears, nearly centered relative to the emission line (the fourth spectrum from the bottom to the top in   Figure\,\ref{LowSpec}, panel B, and  Figure~\ref{fig:HeI}). The appearance of central absorption in \ion{He}{I} is accompanied by the emergence  of \ion{Na}{I} $\lambda 5890$\AA\ and $5896$\AA\ doublet and probably \ion{Mg}{II} $\lambda 4481$\AA, all in the flanks of \ion{He}{I} lines. We believe that the latter forms at the inflated  side-wall of the disc, where the  temperature is T$_{disc}^{out}\gtrsim 5000K$. 
  As the disc continues revealing itself, the blue peak becomes visibly stronger while the red one nearly disappears, which is when the secondary shades solely the edge of the disc moving away from the observer. We can assume  \ion{He}{I} is emanated mostly by the edges of the disc surface, which are still visible in the middle of an eclipse. None of these profile transformations are observed in  Balmer lines.
  We see no need to invoke a disc wind model to explain the presence of \ion{He}{I} line in the spectrum of RW\,Tri at the minimum of an eclipse. At a depth of eclipse, the profile can be formed by two Gaussians from disc extremes still visible on either side of the secondary, while the central parts of the disc, where the absorption forms,  are blocked. However, the situation with other emission lines is different, and they are better accessible by higher resolution spectroscopy.

 \begin{table}
   \caption{The evolution of some emission line parameters during way out of the eclipse.} 
    \label{LineParEcl}
     \centering
     \begin{tabular}{ccccccc}
   \hline \hline
       Orbilal &  I$_{\mathrm{H}\alpha}$ &  EW$_{\mathrm{H}\alpha}$  &  I$_{\ion{He}{I}}$  & EW$_{\ion{He}{I}}$        & I$_{\ion{He}{II}}$ &  EW$_{\ion{He}{II}}$  \\
       phase &                   &  \AA              &   4471\AA        & \AA                    &      4686 \AA  & \AA   \\ \hline
     0.993   &      1.33         &  -75              &   0.31           &       -12.4            &     0.28       &  -9.2     \\
     0.006   &      1.26         &  -72              &   0.35           &       -11.0            &     0.31       &  -10.2     \\ 
     0.019   &      1.39         &  -56              &   0.30           &       -6.4             &     0.52       &  -10.9    \\ 
     0.031   &      1.45         &  -42              &   0.47$^{*}$     &       -3.7             &     0.78       &  -10.6           \\ 
     0.043   &      1.55         &  -30              &   0.40$^{*}$     &       -1.2             &     0.80       &  -8.6      \\
     0.056   &      1.65         &  -27              &   0.39$^{*}$    &       -0.5             &      1.00       &  -6.2      \\  \hline

     \end{tabular}
     \begin{tabular}{l}
   Peak intensity (I) without continuum in  $\times10^{-14}$ ergs cm$^{-2}$ s$^{-1}$  \AA$^{-1}$;\\
    $^*$ - double-peaked;
     \end{tabular}
    
 \end{table}

\subsection{High-resolution spectroscopy}

 The high-resolution Echelle spectra leave no doubt about the two-component structure of the Balmer emission lines
 (Figure~\ref{fig:Ha}).  The trailed spectrum of H$\alpha$ (see Figure~\ref{DopTom}, top-left) clearly shows narrow,
 low-velocity and wide, higher-velocity components, which are approximately in anti-phase. It is very similar to what was
 recently reported by \citet{2017MNRAS.470.1960H} for two other NL systems RW Sex and 1RXS J064434.5+334451. \citet{2017MNRAS.470.1960H} used the same
 technique to separate two (narrow and wide) components of H$\alpha$ profile (Figure~\ref{fig:Ha}). 
In a simple rendition of components, they are Gaussians, characterized by the peak intensity $I$, 
the full width at half-maximum
(FWHM), and their radial velocities depend on the orbital phase of
 \begin{equation}
    \upsilon = \gamma + A \sin[2\pi(\varphi-\varphi_0)]
 \end{equation}
 The parameters  of both Gaussian components of the H$\alpha$ emission core are presented in  Table~\ref{LinePar}.
 \begin{table}
    \caption{Parameters of the Gaussian components of the H$\alpha$ emission line.}
     \centering
     \begin{tabular}{cccccc}
   \hline \hline
        Emission  & A$\equiv$ V $\sin$(i) &  V  & I/I$_c$ & FWHM        & $\varphi_0$ \\
        component &  km s$^{-1}$ & km s$^{-1}$ &             &  km s$^{-1}$ & (phase)  \\ \hline
        Narrow    &   134.8      &   138.3     &  0.35       &  173         &   0.06                \\
        Wide      &   300(50)    &   310(50)   &    0.71     &  1042        &   0.40          \\ \hline
     \end{tabular}
  
     \label{LinePar}
 \end{table}
 \begin{figure*}
\setlength{\unitlength}{1mm}
\resizebox{15.cm}{!}{
\begin{picture}(130,153)(0,0)
\put(0,0) {\includegraphics[width=11cm,angle=0,bb =60 70 550 750, clip]{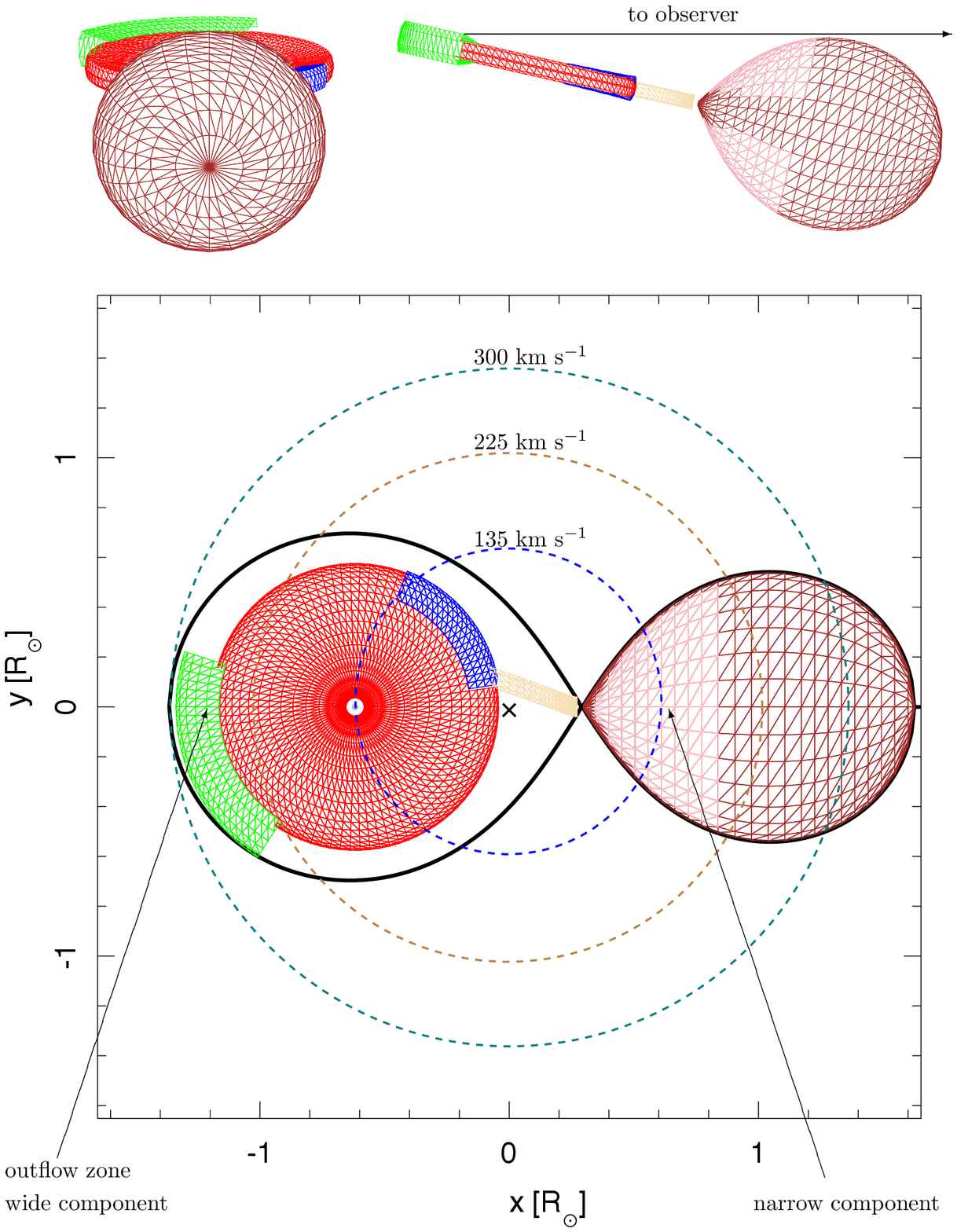}}

\end{picture}}
    \caption{Bottom panel: The geometry of RW~Tri. The dashed circles mark orbits of the primary, the secondary, the disc overflowing regions. The corresponding velocities in km s$^{-1}$ are given. The cross is the center of mass of the system.
    The scale of axes is given in the solar radius.
    Top panel: An observer view on the system in the eclipse in the face and in the profile.}
    \label{Fig:Model}
\end{figure*}

It is obvious that the Gaussian description of components is a simplification. However, it allows us to remove the narrow Gaussian line from the H$\alpha$ emission line profile to probe the origin of both its components. From the trailed spectra, we can infer that the intensity of the narrow component  depends on the orbital phase. The maximum of this component intensity is located at the orbital phase of $\varphi\approx 0.6$ and decreases as the system going to the eclipse. The wide component is complex, with variable peak intensity and FWHM.  As a result, its radial velocity amplitude 
 is determined with large uncertainties.  
 
 \subsection{Doppler tomography}
  We use the Doppler tomography technique \citep{1988MNRAS.235..269M,2001LNP...573....1M} to probe the accretion flow structure in RW\,Tri.
Briefly, the Doppler tomography technique uses the information encoded in spectral line profiles taken at different orbital phases to calculate a distribution of emission over the binary. Doppler tomography provides a quantitative mapping of the optically thin line forming regions in velocity space.  The maximum entropy method  implementation developed by  \citep{1998astro.ph..6141S}\footnote{https://wwwmpa.mpa-garching.mpg.de/~henk/pub/dopmap/}  was used to generated Doppler maps.

 The Doppler tomography of the H$\alpha$ line based on  Echelle spectra is presented in Figure~\ref{DopTom}, top-right. The orbital zero-phase was determined from the photometric data (eclipses), and the secondary Roche lobe size, the accretion stream trajectory, and the maximum size of the disc were derived from the system  parameters determined above. The Doppler map of the entire line, in which the narrow component is quite dominant by its intensity and concentration, 
brilliantly confirms the proposition that the narrow component is formed close to the face-on of the secondary nearly the  L$_1$ point of the system. The residual trailed spectra (observed and reconstructed)  after the narrow component was removed  shown in Figure ~\ref{DopTom}, bottom-left, and center panels. The corresponding Doppler tomogram is presented in  Figure ~\ref{DopTom}, bottom-right. The circle in the Doppler tomogram marks the truncation radius of the disc.
 The residual trailed spectra (observed and reconstructed)  after the  narrow component was removed are shown in Figure ~\ref{DopTom}, bottom-left and -center panels. The corresponding Doppler tomogram is presented in  Figure ~\ref{DopTom}, bottom-right. The circle in the Doppler tomogram marks the truncation radius of the disc.
 This map looks like a horseshoe with two bright extended regions located at 
[V$_x\approx -100$ km s$^{-1}$, V$_y \approx -200$ km s$^{-1}$] and [V$_x\approx 150$ km s$^{-1}$, V$_y \approx 0$ km s$^{-1}$].  The latter is somewhat unusual. 
The velocity map is probably affected by artifacts created by narrow component inaccurate removal; however, this extended spot is certainly not related to the accretion disc. 
  The wide component radial velocity and its reflection in the velocity map attest that it has a lower value than  the minimal Keplerian velocity $ \upsilon_{\mathrm {min}}\sin (i) = 470$ km s$^{-1}$ in the accretion disc may have.
 Presumably, this component of the emission line is not formed in the accretion disc either. It is a common feature detected in high-resolution spectra of the NLs  RW Sex and 1RXS J064434.5+334451. Similar Doppler tomograms were reported for another long period NLs such as AC~Cnc and V363~Aur \citep{1994ApJS...93..519K, 2004MNRAS.353.1135T}. 
 
  As a result of the systematic study of NLs  affiliated to UX\,UMa or RW\,Sex type by high-resolution spectroscopy, we drew a consistent picture, according to which the Balmer emission lines arise from two separate sources. Doppler maps consistently reproduce a pattern very similar to Figure ~\ref{DopTom}, and it permits us to discuss a common phenomenological model for this type of CVs. 

\begin{table*}
	\centering
	\caption{List of NLs (P$_{orb}\gtrapprox4h$) and their principal characteristics}
	\label{tab:Nls}
	\begin{tabular}{lcccccccccc} 
		\hline
		Object & Orbital& Dist. & G band  &incl. & abs. & Narrow  & Wide  & Spiral arms$^\ddagger$ & \ion{He} {II}/H$\beta^{\ddagger\ddagger}$  & Ref. \\
	          & period  &    &    &angle  & presence &  km s$^{-1}$& km s$^{-1}$ & comp. & I$_{\mathrm{He}}$/I$_{\mathrm{H}\beta}$ & \\ 
	            & days  & ps  & mag &degree  &  & $V_{obs}$/$V$ & $V_{obs}$/ $V$ & presence &  & \\ 	\hline
        EC21178-5417 & 0.15452 & 537(9)    & 13.73  & 83(7)     & N        &    -     &     -      &    Y     & 2.0   &  1  \\
        LX Ser       &  0.1584 &  492(10)  &  14.96 &  75(2)    & N        &    -      &     -      &  -      & 0.6  &  2  \\
	   	BG Tri       & 0.15845 & 337(8)    & 11.87  & 21(3)     & Y        & 78/180  & 121/300    &    N    &   weak      &  3  \\
	   	HS 0139+0559 &0.16920  & 1304(102) & 15.07  & n. ecl.   & Y        &   -      &    -       &  -       & N     & 4   \\
	   	IPHAS J210205+471018 
	   	             & 0.1776  & 740(20)   & 15.49   & n. ecl.  & Y        & $175^*$  &    -       &  -       & weak    & 5 \\
	    SDSS 1006+23 &0.18591  &  796(60)  & 18.25  & 81.3(8)   & N        &   -      &    -       &  Y       & 0.5 & 6    \\
	   	IX Vel       &0.19293  & 90.5(2)   & 9.32   & 57(2)     & Y        &   -      &    -       &  Y       & 0.7 & 7  \\
		UX UMa       &0.19667  & 297(2)    & 12.98  & 71        & N        &   -      &    -       &  -       & 0.2 & 8 \\
		V345 Pav     &0.19810  &           & 13.39  & ecl.      & N        &     -    &    -       &  -       & 0.9 & 9 \\
	    V825 Her     &0.20600  & 1078(30)  & 14.12  & n. ecl.   & Y        &   -        &   -      &  -       & -    & 10   \\
	    V3885 Sgr    &0.20714  & 132(1)    & 10.25  & n.ec.     & Y        &   -      &  -         &  Y       & 1.6 & 11   \\
	    2MASS 2256+59&0.22860  &  494(5)   & 14.53  &  78.8(1)  & N        &  -       &  -         &  -       & -    &  12  \\
	    V347 Pup     &0.23119  & 296(1)    & 13.42  & 84.0(2.3) & N        &  -       &  -         &  N       & 0.5 &  13 \\
		RW Tri       &0.23188  &  315(5)   & 13.35  & 77.2(5)   & N        & 132/136  & 300/310    &  N       & 0.5 & 14\\
		DO Leo       &0.23452  & 1463(117) & 16.85  & ecl.      & N        & -        & -          &  -       & 0.2 & 15 \\
		RW Sex       &0.24507  &  230(5)   & 10.63  & 34(6)     & Y        & 50/90    & 165/294    &   N      & weak    & 16 \\
		CM Phe       &0.26890  &  315(4)   & 15.49  & w. ecl.   & N        & -        & -          & -        & -    & 17   \\
		RXJ\,0644+33 &0.26937  & 476(7)    & 13.39  & 74(3)     & N        & 69/71    & 297/309    & N        & 1.0    & 16,18 \\
		AC Cnc      &0.30048  &  673(27)   &  14.26 &  76.3     & N        &  Y       &  Y         &  N       & 0.3    & 19 \\
		V363 Aur     &0.32124  & 486(5)    & 14.13  &  70(2)    & N        &  Y       &  Y         &  N       & 1.3    & 19   \\
		\hline
		\hline
	\end{tabular}
	\begin{tabular}{l}
	1 - \citet{2020MNRAS.491..344R, 2020MNRAS.tmp..637K}; 2 - \citet{1981ApJ...244..259Y}; 3 - Hernandez al. (in preparation); 4 - \citet{2005AA...443..995A}; \\ 5 - \citet{2018ApJ...857...80G}
	6 - \citet{2009AA...507..929S}; 7 - \citet{1983MNRAS.204P..35W, 2007ApJ...662.1204L}; 8 - \citet{2011MNRAS.410..963N}; 9 - \citet{1992MNRAS.258..285B}; \\ 10 - \citet{2005PASP..117.1223R}; 
	11 - \citet{1985AA...151..157H}; 	\citet{2005MNRAS.363..285H}; 12 - \citet{2015AA...584A..40K}; 13 - \citet{2005MNRAS.357..881T}; \\ 14 - this paper; 15 - \citet{1990PASP..102..558A}; 
	16 -  \citet{2017MNRAS.470.1960H} 17 - \citet{1998PASP..110..906H}; \citet{2002MNRAS.335...44W}; \\  18 - \citet{2017MNRAS.464..104H};   19-\citet{2004MNRAS.353.1135T}; \\
	$^\ddagger$ - spiral arm model is proposed in cited references; \\
		$^{\ddagger\ddagger}$ - $\mathrm{I} = \mathrm{I}_{\mathrm{peak}} - \mathrm{I}_{\mathrm{continuum}}$ - rough estimation from published spectra; \\
	n. ecl. - absent of eclipse in the light curve; w. ecl. - low amplitude eclipse is presented. \\
	`-' - absent the data; `Y' - presence of characteristic; `N' - absent of characteristic; `weak' - weak \ion{He} {II} $\lambda 4868$\AA\ in a spectrum with Balmer  adsorptions. \\
	* obtained from a limited, median resolution data
	
	\end{tabular}
\end{table*}

\section{DISCUSSION}
\label{sec:discus}

Among a few  models invoked  to explain  single-peaked emission line profiles  NLs  the disc wind model 
\citep[and references therein]{1995MNRAS.273..225K, 1996Natur.382..789M,  2005ASPC..330..103P, 2010ApJ...719.1932N,
2015MNRAS.450.3331M} prevails. In a  recent study, \citet{2015MNRAS.450.3331M}  explored whether the disc wind models influence 
NLs spectra in the UV/optical wavebands. They applied a standard disc wind model to RW~Tri and found that the model successfully
reproduces the UV spectra of CVs and also leaves a significant imprint on the optical spectrum.  The wind can participate in the
formation of the Balmer and He lines. However, in most cases, no single-peaked line is formed except the narrowing of the
double-peaked emission lines.
Authors also note that the slowly accelerating wind models produce
narrower emission lines, and in order to generate strong optical wind signatures, very high densities ($n_e\sim 10^{13}-10^{14}$
cm$^{-3}$) at the base of the wind are needed. Another problem of these models is that they require considerably stronger
\ion{He}{ii} features than is observed.

Our new high-resolution optical spectroscopy of the eclipsing NL system RW~Tri shows 
a single-peaked but two-component structure of the Balmer lines   and their phase-dependent behavior similar to other NLs  \citep[e.g.,
RW Sex and 1RXS J064434.5+334451, which are populating the same orbital period range,][]{2017MNRAS.470.1960H}. Emission from
\ion{He}{II} is present, but it is  weaker than required by the wind model.   \citet{2017MNRAS.470.1960H} reviewed various
proposed models to explain emission line profiles in NLs and concluded that their behavior in high-mass transfer CVs is
associated with the matter escaping from the accretion disc in the orbital plane, and forming an extended low-velocity region
(hereafter called an outflow zone)  opposite to  hot spot side of the disc. This region is a possible source of the
wide component of the Balmer line profiles in RW~Tri,  RW Sex and 1RXS J064434.5+334451, and some other long-period NLs with
similar Balmer line profiles.  

A list of such NLs is presented in  Table~\ref{tab:Nls}. We selected only systems with known
orbital periods longer than $\approx4$h because a significant fraction of the population of NLs in the 3-4 hours period range
are SW Sex systems \citep{2007MNRAS.377.1747R}. In the latter, the hot spot probably plays a dominant role in the formation of
emission lines \citep{2014AJ....147...68T}. NL-systems with longer orbital periods show distinct characteristics compared to  SW
Sex systems. High-resolution spectroscopy of RW\,Tri, BG\,Tri, RW\,Sex, 1RXS J064434.5+334451, AC Cnc, and V363 Aur exhibit typical
two-component structure of the Balmer lines clearly illustrated by trailed spectra (see references in  Table~\ref{tab:Nls}). 
Here, we put together main spectral characteristics of  NLs including the presence or absence of Balmer absorptions, available
radial velocities of components of the H$\alpha$ emission line, evidence of spiral arms, and presence of \ion{He}{II}
$\lambda$4686\AA\ line or a ratio of peak intensities  I$_{\mathrm{He}}$/I$_{\mathrm{H}\beta}$. Only  low inclination systems exhibit 
Balmer adsorptions. The majority of objects do not exhibit strong \ion{He}{II} emissions, which, as we believe, could be an indication of  the presence 
of  the disc wind. In a few cases, the Doppler maps of Balmer lines
were interpreted as evidence of  spiral arms. 

In Figure~\ref{Fig:Model}, we  reproduce the geometry and an observer view of the system during the eclipse and mark by arrows
both regions, which are responsible for emission line formation according to our interpretation. The system parameters of
RW~Tri, obtained from our optical light curve fit, were used. The narrow, low-velocity component of Balmer emission lines forms 
at the  heated face of the secondary because its radial velocity (138.3 km s$^{-1}$) is about twice larger than the orbital
velocity of the L$_1$ point ($\upsilon_{\mathrm{L}_1} = 64.7$~km~s$^{-1}$). Meanwhile, the radial velocity of the secondary (center of
mass) reaches 225\,km~s$^{-1}$.  The intensity of the narrow component is phase-dependent,  as shown
in the trailed spectra (Figure~\ref{DopTom}, top-left panel). The narrow component is brighter at the orbital phase $\varphi\sim0.6$
and practically disappears around the eclipse. The fact that in  NLs the disc is capable of irradiating the face of the secondary
means,  it also easily excites the gas in the outflow region to produce emission lines. Meanwhile, in dwarf novae, we seldom
detect emission lines from the L$_1$ point, because the optically thin disc can not provide sufficient energetic photons. Hence,
we do not see the outflow region either.  Besides, in dwarf novae, the accretion rate is lower, and the outflow rate is undoubtedly lower.

The wide component of Balmer lines in our model corresponds to the outflow zone located in the opposite to the standard hot spot
position and is always visible in all orbital phases, including the eclipse (see the top panels of Figure~\ref{Fig:Model}). 
On average it has a semi-amplitude of the orbital velocity of $\approx300$~km s$^{-1}$, but naturally shows large velocity
dispersion since the line forming particles have diverse velocities in the non-Keplerian outflow zone. 
Examples of hydrodynamic simulations detailing disc outflow in the orbital plane around this region 
were presented by \citet{2007ARep...51..836S, 2010MmSAI..81..187B, 2017ApJ...841...29J,2019INASR...3..225K}. 
Interestingly this outflow region is confined to the Roche lobe of the white dwarf, and the maximum orbital
velocity of the Roche lobe extreme is also $\sim$ 300~km s$^{-1}$.  If, in contrast, the wide component forms at the base of a
disc wind, it is expected to have a radial velocity compatible with the white dwarf, or more than twice smaller. 
However, the measured velocities of wide and narrow components, when it was possible to separate them clearly, show that the former's semi-amplitude is always higher (see  Table~\ref{tab:Nls}), while in systems with $q <1.0$ should be the opposite. The radial velocity of the wide component in all NLs after correction for the inclination angle (notwithstanding the uncertainties)
is regularly around $\sim$300~km s$^{-1}$. This value is too high for a wind originating from the hot, inner parts of the accretion disc,
but it is reasonable if the wide component formation area is located in the orbital plane beyond the disc.

However, most models of the outflow of matter from accretion discs are associated with possible wind from the accretion disc. We agree that the wind  exists because it allows us to explain common characteristics of the velocity field produced by emission lines in a group of similar objects in the ultraviolet.  Nonetheless, in high mass transfer systems with thick accretion discs, the impact of the mass transfer stream with the disc creates some specific conditions different from CVs with thin discs.   Observations of NLs in UV- domain \citep{1997MNRAS.290L..23M} show a significant dip of the UV flux prior to the eclipse.  In the case of RW\,Tri, this dip is centered at the orbital phase $\varphi \approx 0.7-0.8$, or when the binary is viewed from the side. At this particular angle, neither the hot spot nor the outflow zone, as we identify it in Figure\,\ref{Fig:Model}, is directly blocking the inner disc to produce a dip. However, an alternative solution for the dip is  predicted in the calculation of \citet[see fig.6 therein]{2007ARep...51..836S}. The ejected material through the  L$_3$ Lagrangian point (coinciding with the outflow region in our model) spreads further and  can absorb the observed light curve, significantly stronger in UV than in other domains. We consider this notion as an additional argument in support of the idea that the outflow zone exists.

\section{CONCLUSIONS}
\label{sec:Concl}

We conducted low  and high-resolution spectroscopy of the eclipsing nova-like system RW\,Tri. We also obtained photometric
observations of the object as well as gathered all available AAVSO data together with data from the literature. 
Using our specialized code developed to model light curves of binary systems along with precise GAIA distance based on the system's spectral and photometric characteristics, we deduced parameters of the object and their uncertainties.
We consider that our estimates improve similar evaluations in the past.

The more important outcome of our study addresses the problem related to the structure and the origin of emission
lines. RW\,Tri is an exemplar of NLs  in which two components of Balmer emission lines are clearly separated 
(see Table\,\ref{tab:Nls}). While the origin of the narrow component easily discernible, the other component is still very much debatable. 
\citet{2017MNRAS.470.1960H} proposed  the accretion disk spills out at the opposed to the secondary side.  
We argue that this dominant component of Balmer emission lines is produced by the gas outflow zone marked in Figure\,\ref{Fig:Model}.   This outflow zone is not eclipsed by the secondary, has the right orbital velocity, and a large dispersion in the emitting particle velocities. It probably forms a material escaping the system through L$_3$ vicinity and which is responsible for some observed phenomenons in the system. We believe that the disc has sufficient flux to ignite emission lines from the outflow zone and from the secondary star surface adjacent to the  L$_1$ point. 

We do not question the existence of accretion disc wind and its presence in RW Tri; however, the wind does no form any significant emission in optical hydrogen or helium lines. The wind affects the formation of the UV spectrum of the system, but the wind models need unrealistically high densities and intense high energy fluxes to create single-peaked emission lines in the optical domain. Another problem of the wind models is the Balmer lines' two-component structure and a relatively high orbital velocity of its wide component Also, wind models predict the strong \ion{He}{II} $\lambda 4686$\AA\ line, similar to that observed, for example, in V Sge-type stars \citep{1998PASP..110..276S}.  Meanwhile, \ion{He}{ii} is often weak or almost non-existent in a number of NLs (see Table\,\ref{tab:Nls}).

We consider SW\,Sex-type stars  (3h $\lesssim \mathrm{P}_{\mathrm{orb}} \lesssim 4$h) as an evolutionary descendant of NLs discussed here. A shrinking orbit results in some qualitative changes in density/temperature of the hot spot, converting it into a dominant source of emission lines, while the irradiation of the secondary and outflow region lose their significance. We should also point out that the outflow zone does not disappear, but it becomes a source of absorption features instead \citep{2014AJ....147...68T}. 
Long-period NLs are few and poorly studied, the number of bright sources available for the high-resolution spectroscopy with small size telescopes ($\lesssim $2m) is limited. We believe that future high-resolution spectroscopy at larger telescopes will confirm our suggestion about the origin of the emission line components and the general idea of the similarity of accretion flow structure.  

\section*{Acknowledgement}

This work is based upon observations carried out at the OAN SPM, Baja California, M\'{e}xico. SZ  and GT acknowledge PAPIIT grants IN108316, IN102120, IN110619 and CONACyT grant 166376. SZ and VN acknowledge the financial support from the visitor and mobility program of the Finnish Centre for Astronomy with ESO (FINCA), funded by the Academy of Finland grant nr 306531. MSH acknowledges the Fellowship for National PhD from ANID, grant number 21170070. The research of MW was supported by the project Progress Q47 Physics of the Charles University  in Prague.
We acknowledge with thanks the variable star observations from the AAVSO International Database contributed by observers worldwide and used in this research. We thank the daytime and night support staff at the OAN-SPM for facilitating and helping obtain our observations.
 We are grateful to the anonymous referee for the useful valuable suggestions that allowed us to improve the manuscript.

\section*{Data Availability}

The data underlying this article will be shared on reasonable request to the corresponding author.



\bibliographystyle{mnras}
\bibliography{subebekovaRWTri}

\begin{thebibliography}{}
\makeatletter
\relax
\def\mn@urlcharsother{\let\do\@makeother \do\$\do\&\do\#\do\^\do\_\do\%\do\~}
\def\mn@doi{\begingroup\mn@urlcharsother \@ifnextchar [ {\mn@doi@}
  {\mn@doi@[]}}
\def\mn@doi@[#1]#2{\def\@tempa{#1}\ifx\@tempa\@empty \href
  {http://dx.doi.org/#2} {doi:#2}\else \href {http://dx.doi.org/#2} {#1}\fi
  \endgroup}
\def\mn@eprint#1#2{\mn@eprint@#1:#2::\@nil}
\def\mn@eprint@arXiv#1{\href {http://arxiv.org/abs/#1} {{\tt arXiv:#1}}}
\def\mn@eprint@dblp#1{\href {http://dblp.uni-trier.de/rec/bibtex/#1.xml}
  {dblp:#1}}
\def\mn@eprint@#1:#2:#3:#4\@nil{\def\@tempa {#1}\def\@tempb {#2}\def\@tempc
  {#3}\ifx \@tempc \@empty \let \@tempc \@tempb \let \@tempb \@tempa \fi \ifx
  \@tempb \@empty \def\@tempb {arXiv}\fi \@ifundefined
  {mn@eprint@\@tempb}{\@tempb:\@tempc}{\expandafter \expandafter \csname
  mn@eprint@\@tempb\endcsname \expandafter{\@tempc}}}

\bibitem[\protect\citeauthoryear{{Abbott}, {Shafter}, {Wood}, {Tomaney}  \&
  {Haswell}}{{Abbott} et~al.}{1990}]{1990PASP..102..558A}
{Abbott} T.~M.~C.,  {Shafter} A.~W.,  {Wood} J.~H.,  {Tomaney} A.~B.,
  {Haswell} C.~A.,  1990, \mn@doi [\pasp] {10.1086/132669}, \href
  {https://ui.adsabs.harvard.edu/abs/1990PASP..102..558A} {102, 558}

\bibitem[\protect\citeauthoryear{{Africano}, {Nather}, {Patterson}, {Robinson}
  \& {Warner}}{{Africano} et~al.}{1978}]{1978PASP...90..568A}
{Africano} J.~L.,  {Nather} R.~E.,  {Patterson} J.,  {Robinson} E.~L.,
  {Warner} B.,  1978, \mn@doi [\pasp] {10.1086/130387}, \href
  {https://ui.adsabs.harvard.edu/abs/1978PASP...90..568A} {90, 568}

\bibitem[\protect\citeauthoryear{{Applegate}}{{Applegate}}{1992}]{1992ApJ...385..621A}
{Applegate} J.~H.,  1992, \mn@doi [\apj] {10.1086/170967}, \href
  {https://ui.adsabs.harvard.edu/abs/1992ApJ...385..621A} {385, 621}

\bibitem[\protect\citeauthoryear{{Aungwerojwit} et~al.,}{{Aungwerojwit}
  et~al.}{2005}]{2005AA...443..995A}
{Aungwerojwit} A.,  et~al., 2005, \mn@doi [\aap] {10.1051/0004-6361:20042610},
  \href {https://ui.adsabs.harvard.edu/abs/2005A&A...443..995A} {443, 995}

\bibitem[\protect\citeauthoryear{{Bisikalo} \& {Kononov}}{{Bisikalo} \&
  {Kononov}}{2010}]{2010MmSAI..81..187B}
{Bisikalo} D.~V.,  {Kononov} D.~A.,  2010, \memsai, \href
  {https://ui.adsabs.harvard.edu/abs/2010MmSAI..81..187B} {81, 187}

\bibitem[\protect\citeauthoryear{{Boyd}}{{Boyd}}{2012}]{2012JAVSO..40..295B}
{Boyd} D.,  2012, Journal of the American Association of Variable Star
  Observers (JAAVSO), \href
  {https://ui.adsabs.harvard.edu/abs/2012JAVSO..40..295B} {40, 295}

\bibitem[\protect\citeauthoryear{{Buckley}, {O'Donoghue}, {Kilkenny}, {Stobie}
  \& {Remillard}}{{Buckley} et~al.}{1992}]{1992MNRAS.258..285B}
{Buckley} D.~A.~H.,  {O'Donoghue} D.,  {Kilkenny} D.,  {Stobie} R.~S.,
  {Remillard} R.~A.,  1992, \mn@doi [\mnras] {10.1093/mnras/258.2.285}, \href
  {https://ui.adsabs.harvard.edu/abs/1992MNRAS.258..285B} {258, 285}

\bibitem[\protect\citeauthoryear{{Claret}, {Hauschildt}  \& {Witte}}{{Claret}
  et~al.}{2012}]{2012A&A...546A..14C}
{Claret} A.,  {Hauschildt} P.~H.,   {Witte} S.,  2012, \mn@doi [\aap]
  {10.1051/0004-6361/201219849}, \href
  {https://ui.adsabs.harvard.edu/abs/2012A&A...546A..14C} {546, A14}

\bibitem[\protect\citeauthoryear{{Dhillon}, {Marsh}  \& {Jones}}{{Dhillon}
  et~al.}{1997}]{1997MNRAS.291..694D}
{Dhillon} V.~S.,  {Marsh} T.~R.,   {Jones} D.~H.~P.,  1997, \mn@doi [\mnras]
  {10.1093/mnras/291.4.694}, \href
  {https://ui.adsabs.harvard.edu/abs/1997MNRAS.291..694D} {291, 694}

\bibitem[\protect\citeauthoryear{{Dhillon}, {Littlefair}, {Howell}, {Ciardi},
  {Harrop-Allin}  \& {Marsh}}{{Dhillon} et~al.}{2000}]{2000MNRAS.314..826D}
{Dhillon} V.~S.,  {Littlefair} S.~P.,  {Howell} S.~B.,  {Ciardi} D.~R.,
  {Harrop-Allin} M.~K.,   {Marsh} T.~R.,  2000, \mn@doi [\mnras]
  {10.1046/j.1365-8711.2000.03427.x}, \href
  {https://ui.adsabs.harvard.edu/abs/2000MNRAS.314..826D} {314, 826}

\bibitem[\protect\citeauthoryear{{Dmitrienko}}{{Dmitrienko}}{1992}]{1992BCrAO..86...58D}
{Dmitrienko} E.~S.,  1992, Bulletin Crimean Astrophysical Observatory, \href
  {https://ui.adsabs.harvard.edu/abs/1992BCrAO..86...58D} {86, 58}

\bibitem[\protect\citeauthoryear{{Frank} \& {King}}{{Frank} \&
  {King}}{1981}]{1981MNRAS.195..227F}
{Frank} J.,  {King} A.~R.,  1981, \mn@doi [\mnras] {10.1093/mnras/195.2.227},
  \href {https://ui.adsabs.harvard.edu/abs/1981MNRAS.195..227F} {195, 227}

\bibitem[\protect\citeauthoryear{{Guerrero} et~al.,}{{Guerrero}
  et~al.}{2018}]{2018ApJ...857...80G}
{Guerrero} M.~A.,  et~al., 2018, \mn@doi [\apj] {10.3847/1538-4357/aab669},
  \href {https://ui.adsabs.harvard.edu/abs/2018ApJ...857...80G} {857, 80}

\bibitem[\protect\citeauthoryear{{Hartley}, {Murray}, {Drew}  \&
  {Long}}{{Hartley} et~al.}{2005}]{2005MNRAS.363..285H}
{Hartley} L.~E.,  {Murray} J.~R.,  {Drew} J.~E.,   {Long} K.~S.,  2005, \mn@doi
  [\mnras] {10.1111/j.1365-2966.2005.09447.x}, \href
  {https://ui.adsabs.harvard.edu/abs/2005MNRAS.363..285H} {363, 285}

\bibitem[\protect\citeauthoryear{{Haug} \& {Drechsel}}{{Haug} \&
  {Drechsel}}{1985}]{1985AA...151..157H}
{Haug} K.,  {Drechsel} H.,  1985, \aap, \href
  {https://ui.adsabs.harvard.edu/abs/1985A&A...151..157H} {151, 157}

\bibitem[\protect\citeauthoryear{{Hellier} \& {Robinson}}{{Hellier} \&
  {Robinson}}{1994}]{1994ApJ...431L.107H}
{Hellier} C.,  {Robinson} E.~L.,  1994, \mn@doi [\apjl] {10.1086/187484}, \href
  {https://ui.adsabs.harvard.edu/abs/1994ApJ...431L.107H} {431, L107}

\bibitem[\protect\citeauthoryear{{Hensler}}{{Hensler}}{1982}]{1982A&A...114..319H}
{Hensler} G.,  1982, \aap, \href
  {https://ui.adsabs.harvard.edu/abs/1982A&A...114..319H} {114, 319}

\bibitem[\protect\citeauthoryear{{Hern{\'a}ndez Santisteban},
  {Echevarr{\'\i}a}, {Michel}  \& {Costero}}{{Hern{\'a}ndez Santisteban}
  et~al.}{2017}]{2017MNRAS.464..104H}
{Hern{\'a}ndez Santisteban} J.~V.,  {Echevarr{\'\i}a} J.,  {Michel} R.,
  {Costero} R.,  2017, \mn@doi [\mnras] {10.1093/mnras/stw2282}, \href
  {https://ui.adsabs.harvard.edu/abs/2017MNRAS.464..104H} {464, 104}

\bibitem[\protect\citeauthoryear{{Hernandez}, {Zharikov}, {Neustroev}  \&
  {Tovmassian}}{{Hernandez} et~al.}{2017}]{2017MNRAS.470.1960H}
{Hernandez} M.~S.,  {Zharikov} S.,  {Neustroev} V.,   {Tovmassian} G.,  2017,
  \mn@doi [\mnras] {10.1093/mnras/stx1341}, \href
  {https://ui.adsabs.harvard.edu/abs/2017MNRAS.470.1960H} {470, 1960}

\bibitem[\protect\citeauthoryear{{Hoard} \& {Wachter}}{{Hoard} \&
  {Wachter}}{1998}]{1998PASP..110..906H}
{Hoard} D.~W.,  {Wachter} S.,  1998, \mn@doi [\pasp] {10.1086/316212}, \href
  {https://ui.adsabs.harvard.edu/abs/1998PASP..110..906H} {110, 906}

\bibitem[\protect\citeauthoryear{{Hoard} et~al.,}{{Hoard}
  et~al.}{2014}]{2014ApJ...786...68H}
{Hoard} D.~W.,  et~al., 2014, \mn@doi [\apj] {10.1088/0004-637X/786/1/68},
  \href {https://ui.adsabs.harvard.edu/abs/2014ApJ...786...68H} {786, 68}

\bibitem[\protect\citeauthoryear{{Honeycutt}, {Schlegel}  \&
  {Kaitchuck}}{{Honeycutt} et~al.}{1986}]{1986ApJ...302..388H}
{Honeycutt} R.~K.,  {Schlegel} E.~M.,   {Kaitchuck} R.~H.,  1986, \mn@doi
  [\apj] {10.1086/163997}, \href
  {https://ui.adsabs.harvard.edu/abs/1986ApJ...302..388H} {302, 388}

\bibitem[\protect\citeauthoryear{{Horne} \& {Stiening}}{{Horne} \&
  {Stiening}}{1985}]{1985MNRAS.216..933H}
{Horne} K.,  {Stiening} R.~F.,  1985, \mn@doi [\mnras]
  {10.1093/mnras/216.4.933}, \href
  {https://ui.adsabs.harvard.edu/abs/1985MNRAS.216..933H} {216, 933}

\bibitem[\protect\citeauthoryear{{Horne}, {Lanning}  \& {Gomer}}{{Horne}
  et~al.}{1982}]{1982ApJ...252..681H}
{Horne} K.,  {Lanning} H.~H.,   {Gomer} R.~H.,  1982, \mn@doi [\apj]
  {10.1086/159594}, \href
  {https://ui.adsabs.harvard.edu/abs/1982ApJ...252..681H} {252, 681}

\bibitem[\protect\citeauthoryear{{Ju}, {Stone}  \& {Zhu}}{{Ju}
  et~al.}{2017}]{2017ApJ...841...29J}
{Ju} W.,  {Stone} J.~M.,   {Zhu} Z.,  2017, \mn@doi [\apj]
  {10.3847/1538-4357/aa705d}, \href
  {https://ui.adsabs.harvard.edu/abs/2017ApJ...841...29J} {841, 29}

\bibitem[\protect\citeauthoryear{{Kaitchuck}, {Honeycutt}  \&
  {Schlegel}}{{Kaitchuck} et~al.}{1983}]{1983ApJ...267..239K}
{Kaitchuck} R.~H.,  {Honeycutt} R.~K.,   {Schlegel} E.~M.,  1983, \mn@doi
  [\apj] {10.1086/160863}, \href
  {https://ui.adsabs.harvard.edu/abs/1983ApJ...267..239K} {267, 239}

\bibitem[\protect\citeauthoryear{{Kaitchuck}, {Schlegel}, {Honeycutt}, {Horne},
  {Marsh}, {White}  \& {Mansperger}}{{Kaitchuck}
  et~al.}{1994}]{1994ApJS...93..519K}
{Kaitchuck} R.~H.,  {Schlegel} E.~M.,  {Honeycutt} R.~K.,  {Horne} K.,  {Marsh}
  T.~R.,  {White} J.~C. I.,   {Mansperger} C.~S.,  1994, \mn@doi [\apjs]
  {10.1086/192065}, \href
  {https://ui.adsabs.harvard.edu/abs/1994ApJS...93..519K} {93, 519}

\bibitem[\protect\citeauthoryear{{Kaygorodov}}{{Kaygorodov}}{2019}]{2019INASR...3..225K}
{Kaygorodov} P.~V.,  2019, \mn@doi [INASAN Science Reports]
  {10.26087/INASAN.2019.3.1.035}, \href
  {https://ui.adsabs.harvard.edu/abs/2019INASR...3..225K} {3, 225}

\bibitem[\protect\citeauthoryear{{Khangale}, {Woudt}, {Potter}, {Warner},
  {Kilkenny}  \& {van der Heyden}}{{Khangale}
  et~al.}{2020}]{2020MNRAS.tmp..637K}
{Khangale} Z.~N.,  {Woudt} P.~A.,  {Potter} S.~B.,  {Warner} B.,  {Kilkenny}
  D.,   {van der Heyden} K.,  2020, \mn@doi [\mnras] {10.1093/mnras/staa680},
  \href {https://ui.adsabs.harvard.edu/abs/2020MNRAS.tmp..637K} {}

\bibitem[\protect\citeauthoryear{{Kjurkchieva}, {Khruzina}, {Dimitrov},
  {Groebel}, {Ibryamov}  \& {Nikolov}}{{Kjurkchieva}
  et~al.}{2015}]{2015AA...584A..40K}
{Kjurkchieva} D.,  {Khruzina} T.,  {Dimitrov} D.,  {Groebel} R.,  {Ibryamov}
  S.,   {Nikolov} G.,  2015, \mn@doi [\aap] {10.1051/0004-6361/201526102},
  \href {https://ui.adsabs.harvard.edu/abs/2015A&A...584A..40K} {584, A40}

\bibitem[\protect\citeauthoryear{{Knigge}, {Woods}  \& {Drew}}{{Knigge}
  et~al.}{1995}]{1995MNRAS.273..225K}
{Knigge} C.,  {Woods} J.~A.,   {Drew} J.~E.,  1995, \mn@doi [\mnras]
  {10.1093/mnras/273.2.225}, \href
  {https://ui.adsabs.harvard.edu/abs/1995MNRAS.273..225K} {273, 225}

\bibitem[\protect\citeauthoryear{{Kunze}, {Speith}  \& {Hessman}}{{Kunze}
  et~al.}{2001}]{2001MNRAS.322..499K}
{Kunze} S.,  {Speith} R.,   {Hessman} F.~V.,  2001, \mn@doi [\mnras]
  {10.1046/j.1365-8711.2001.04057.x}, \href
  {https://ui.adsabs.harvard.edu/abs/2001MNRAS.322..499K} {322, 499}

\bibitem[\protect\citeauthoryear{{Lanza} \& {Rodon{\`o}}}{{Lanza} \&
  {Rodon{\`o}}}{1999}]{1999A&A...349..887L}
{Lanza} A.~F.,  {Rodon{\`o}} M.,  1999, \aap, \href
  {https://ui.adsabs.harvard.edu/abs/1999A&A...349..887L} {349, 887}

\bibitem[\protect\citeauthoryear{{Levine} \& {Chakarabarty}}{{Levine} \&
  {Chakarabarty}}{1995}]{1995Levin}
{Levine} S.,  {Chakarabarty} D.,  1995, IA-UNAM Technical Report MU-94-04

\bibitem[\protect\citeauthoryear{{Li} et~al.,}{{Li}
  et~al.}{2017}]{2017PASJ...69...28L}
{Li} K.,  et~al., 2017, \mn@doi [\pasj] {10.1093/pasj/psw134}, \href
  {https://ui.adsabs.harvard.edu/abs/2017PASJ...69...28L} {69, 28}

\bibitem[\protect\citeauthoryear{{Lin}, {Williams}  \& {Stover}}{{Lin}
  et~al.}{1988}]{1988ApJ...327..234L}
{Lin} D.~N.~C.,  {Williams} R.~E.,   {Stover} R.~J.,  1988, \mn@doi [\apj]
  {10.1086/166185}, \href
  {https://ui.adsabs.harvard.edu/abs/1988ApJ...327..234L} {327, 234}

\bibitem[\protect\citeauthoryear{{Linnell}, {Godon}, {Hubeny}, {Sion}  \&
  {Szkody}}{{Linnell} et~al.}{2007}]{2007ApJ...662.1204L}
{Linnell} A.~P.,  {Godon} P.,  {Hubeny} I.,  {Sion} E.~M.,   {Szkody} P.,
  2007, \mn@doi [\apj] {10.1086/517965}, \href
  {https://ui.adsabs.harvard.edu/abs/2007ApJ...662.1204L} {662, 1204}

\bibitem[\protect\citeauthoryear{{Longmore}, {Lee}, {Allen}  \&
  {Adams}}{{Longmore} et~al.}{1981}]{1981MNRAS.195..825L}
{Longmore} A.~J.,  {Lee} T.~J.,  {Allen} D.~A.,   {Adams} D.~J.,  1981, \mn@doi
  [\mnras] {10.1093/mnras/195.4.825}, \href
  {https://ui.adsabs.harvard.edu/abs/1981MNRAS.195..825L} {195, 825}

\bibitem[\protect\citeauthoryear{{Luri} et~al.,}{{Luri}
  et~al.}{2018}]{2018A&A...616A...9L}
{Luri} X.,  et~al., 2018, \mn@doi [\aap] {10.1051/0004-6361/201832964}, \href
  {https://ui.adsabs.harvard.edu/abs/2018A&A...616A...9L} {616, A9}

\bibitem[\protect\citeauthoryear{{Marsh}}{{Marsh}}{2001}]{2001LNP...573....1M}
{Marsh} T.~R.,  2001, {Doppler Tomography}.
p.~1

\bibitem[\protect\citeauthoryear{{Marsh} \& {Horne}}{{Marsh} \&
  {Horne}}{1988}]{1988MNRAS.235..269M}
{Marsh} T.~R.,  {Horne} K.,  1988, \mn@doi [\mnras] {10.1093/mnras/235.1.269},
  \href {https://ui.adsabs.harvard.edu/abs/1988MNRAS.235..269M} {235, 269}

\bibitem[\protect\citeauthoryear{{Mason}, {Drew}  \& {Knigge}}{{Mason}
  et~al.}{1997}]{1997MNRAS.290L..23M}
{Mason} K.~O.,  {Drew} J.~E.,   {Knigge} C.,  1997, \mn@doi [\mnras]
  {10.1093/mnras/290.2.L23}, \href
  {https://ui.adsabs.harvard.edu/abs/1997MNRAS.290L..23M} {290, L23}

\bibitem[\protect\citeauthoryear{{Matthews}, {Knigge}, {Long}, {Sim}  \&
  {Higginbottom}}{{Matthews} et~al.}{2015}]{2015MNRAS.450.3331M}
{Matthews} J.~H.,  {Knigge} C.,  {Long} K.~S.,  {Sim} S.~A.,   {Higginbottom}
  N.,  2015, \mn@doi [\mnras] {10.1093/mnras/stv867}, \href
  {https://ui.adsabs.harvard.edu/abs/2015MNRAS.450.3331M} {450, 3331}

\bibitem[\protect\citeauthoryear{{Mauche}, {Raymond}, {Buckley}, {Mouchet},
  {Bonnell}, {Sullivan}, {Bonnet-Bidaud}  \& {Bunk}}{{Mauche}
  et~al.}{1994}]{1994ApJ...424..347M}
{Mauche} C.~W.,  {Raymond} J.~C.,  {Buckley} D. A.~H.,  {Mouchet} M.,
  {Bonnell} J.,  {Sullivan} D.~J.,  {Bonnet-Bidaud} J.-M.,   {Bunk} W.~H.,
  1994, \mn@doi [\apj] {10.1086/173894}, \href
  {https://ui.adsabs.harvard.edu/abs/1994ApJ...424..347M} {424, 347}

\bibitem[\protect\citeauthoryear{{McArthur} et~al.,}{{McArthur}
  et~al.}{1999}]{1999ApJ...520L..59M}
{McArthur} B.~E.,  et~al., 1999, \mn@doi [\apjl] {10.1086/312135}, \href
  {https://ui.adsabs.harvard.edu/abs/1999ApJ...520L..59M} {520, L59}

\bibitem[\protect\citeauthoryear{{Meglicki}, {Wickramasinghe}  \&
  {Bicknell}}{{Meglicki} et~al.}{1993}]{1993MNRAS.264..691M}
{Meglicki} Z.,  {Wickramasinghe} D.,   {Bicknell} G.~V.,  1993, \mn@doi
  [\mnras] {10.1093/mnras/264.3.691}, \href
  {https://ui.adsabs.harvard.edu/abs/1993MNRAS.264..691M} {264, 691}

\bibitem[\protect\citeauthoryear{{Murray}}{{Murray}}{1996}]{1996MNRAS.279..402M}
{Murray} J.~R.,  1996, \mn@doi [\mnras] {10.1093/mnras/279.2.402}, \href
  {https://ui.adsabs.harvard.edu/abs/1996MNRAS.279..402M} {279, 402}

\bibitem[\protect\citeauthoryear{{Murray} \& {Chiang}}{{Murray} \&
  {Chiang}}{1996}]{1996Natur.382..789M}
{Murray} N.,  {Chiang} J.,  1996, \mn@doi [\nat] {10.1038/382789a0}, \href
  {https://ui.adsabs.harvard.edu/abs/1996Natur.382..789M} {382, 789}

\bibitem[\protect\citeauthoryear{{Neustroev}, {Suleimanov}, {Borisov},
  {Belyakov}  \& {Shearer}}{{Neustroev} et~al.}{2011}]{2011MNRAS.410..963N}
{Neustroev} V.~V.,  {Suleimanov} V.~F.,  {Borisov} N.~V.,  {Belyakov} K.~V.,
  {Shearer} A.,  2011, \mn@doi [\mnras] {10.1111/j.1365-2966.2010.17495.x},
  \href {https://ui.adsabs.harvard.edu/abs/2011MNRAS.410..963N} {410, 963}

\bibitem[\protect\citeauthoryear{{Noebauer}, {Long}, {Sim}  \&
  {Knigge}}{{Noebauer} et~al.}{2010}]{2010ApJ...719.1932N}
{Noebauer} U.~M.,  {Long} K.~S.,  {Sim} S.~A.,   {Knigge} C.,  2010, \mn@doi
  [\apj] {10.1088/0004-637X/719/2/1932}, \href
  {https://ui.adsabs.harvard.edu/abs/2010ApJ...719.1932N} {719, 1932}

\bibitem[\protect\citeauthoryear{{Paschke} \& {Brat}}{{Paschke} \&
  {Brat}}{2006}]{2006OEJV...23...13P}
{Paschke} A.,  {Brat} L.,  2006, Open European Journal on Variable Stars, \href
  {https://ui.adsabs.harvard.edu/abs/2006OEJV...23...13P} {23, 13}

\bibitem[\protect\citeauthoryear{{Pecaut} \& {Mamajek}}{{Pecaut} \&
  {Mamajek}}{2013}]{2013ApJS..208....9P}
{Pecaut} M.~J.,  {Mamajek} E.~E.,  2013, \mn@doi [\apjs]
  {10.1088/0067-0049/208/1/9}, \href
  {https://ui.adsabs.harvard.edu/abs/2013ApJS..208....9P} {208, 9}

\bibitem[\protect\citeauthoryear{{Polsgrove}, {Wetterer}, {Bloomer}  \&
  {Newton}}{{Polsgrove} et~al.}{2006}]{2006IBVS.5710....1P}
{Polsgrove} D.~E.,  {Wetterer} C.~J.,  {Bloomer} R.~H.,   {Newton} J.~D.,
  2006, Information Bulletin on Variable Stars, \href
  {https://ui.adsabs.harvard.edu/abs/2006IBVS.5710....1P} {5710, 1}

\bibitem[\protect\citeauthoryear{{Poole}, {Mason}, {Ramsay}, {Drew}  \&
  {Smith}}{{Poole} et~al.}{2003}]{2003MNRAS.340..499P}
{Poole} T.,  {Mason} K.~O.,  {Ramsay} G.,  {Drew} J.~E.,   {Smith} R.~C.,
  2003, \mn@doi [\mnras] {10.1046/j.1365-8711.2003.06316.x}, \href
  {https://ui.adsabs.harvard.edu/abs/2003MNRAS.340..499P} {340, 499}

\bibitem[\protect\citeauthoryear{{Proga}}{{Proga}}{2005}]{2005ASPC..330..103P}
{Proga} D.,  2005, {Theory of Outflows in Cataclysmic Variables}.
p.~103

\bibitem[\protect\citeauthoryear{{Protitch}}{{Protitch}}{1938}]{1938AN....266...95P}
{Protitch} M.,  1938, Astronomische Nachrichten, \href
  {https://ui.adsabs.harvard.edu/abs/1938AN....266...95P} {266, 95}

\bibitem[\protect\citeauthoryear{{Puebla}, {Diaz}, {Hillier}  \&
  {Hubeny}}{{Puebla} et~al.}{2011}]{2011ApJ...736...17P}
{Puebla} R.~E.,  {Diaz} M.~P.,  {Hillier} D.~J.,   {Hubeny} I.,  2011, \mn@doi
  [\apj] {10.1088/0004-637X/736/1/17}, \href
  {https://ui.adsabs.harvard.edu/abs/2011ApJ...736...17P} {736, 17}

\bibitem[\protect\citeauthoryear{{Ringwald}, {Chase}  \& {Reynolds}}{{Ringwald}
  et~al.}{2005}]{2005PASP..117.1223R}
{Ringwald} F.~A.,  {Chase} D.~W.,   {Reynolds} D.~S.,  2005, \mn@doi [\pasp]
  {10.1086/491721}, \href
  {https://ui.adsabs.harvard.edu/abs/2005PASP..117.1223R} {117, 1223}

\bibitem[\protect\citeauthoryear{{Robinson}, {Shetrone}  \&
  {Africano}}{{Robinson} et~al.}{1991}]{1991AJ....102.1176R}
{Robinson} E.~L.,  {Shetrone} M.~D.,   {Africano} J.~L.,  1991, \mn@doi [\aj]
  {10.1086/115944}, \href
  {https://ui.adsabs.harvard.edu/abs/1991AJ....102.1176R} {102, 1176}

\bibitem[\protect\citeauthoryear{{Rodr{\'\i}guez-Gil}
  et~al.,}{{Rodr{\'\i}guez-Gil} et~al.}{2007}]{2007MNRAS.377.1747R}
{Rodr{\'\i}guez-Gil} P.,  et~al., 2007, \mn@doi [\mnras]
  {10.1111/j.1365-2966.2007.11743.x}, \href
  {https://ui.adsabs.harvard.edu/abs/2007MNRAS.377.1747R} {377, 1747}

\bibitem[\protect\citeauthoryear{{Ruiz-Carmona}, {Khangale}, {Woudt}  \&
  {Groot}}{{Ruiz-Carmona} et~al.}{2020}]{2020MNRAS.491..344R}
{Ruiz-Carmona} R.,  {Khangale} Z.~N.,  {Woudt} P.~A.,   {Groot} P.~J.,  2020,
  \mn@doi [\mnras] {10.1093/mnras/stz2839}, \href
  {https://ui.adsabs.harvard.edu/abs/2020MNRAS.491..344R} {491, 344}

\bibitem[\protect\citeauthoryear{{Rutten} \& {Dhillon}}{{Rutten} \&
  {Dhillon}}{1992}]{1992A&A...253..139R}
{Rutten} R.~G.~M.,  {Dhillon} V.~S.,  1992, \aap, \href
  {https://ui.adsabs.harvard.edu/abs/1992A&A...253..139R} {253, 139}

\bibitem[\protect\citeauthoryear{{Rutten}, {van Paradijs}  \&
  {Tinbergen}}{{Rutten} et~al.}{1992}]{1992AA...260..213R}
{Rutten} R.~G.~M.,  {van Paradijs} J.,   {Tinbergen} J.,  1992, \aap, \href
  {https://ui.adsabs.harvard.edu/abs/1992A&A...260..213R} {260, 213}

\bibitem[\protect\citeauthoryear{{Smak}}{{Smak}}{1979}]{1979AcA....29..469S}
{Smak} J.,  1979, \actaa, \href
  {https://ui.adsabs.harvard.edu/abs/1979AcA....29..469S} {29, 469}

\bibitem[\protect\citeauthoryear{{Smak}}{{Smak}}{1995}]{1995AcA....45..259S}
{Smak} J.,  1995, \actaa, \href
  {https://ui.adsabs.harvard.edu/abs/1995AcA....45..259S} {45, 259}

\bibitem[\protect\citeauthoryear{{Smak}}{{Smak}}{2019}]{2019AcA....69...79S}
{Smak} J.,  2019, \mn@doi [\actaa] {10.32023/0001-5237/69.1.6}, \href
  {https://ui.adsabs.harvard.edu/abs/2019AcA....69...79S} {69, 79}

\bibitem[\protect\citeauthoryear{{Southworth}, {Hickman}, {Marsh},
  {Rebassa-Mansergas}, {G{\"a}nsicke}, {Copperwheat}  \&
  {Rodr{\'\i}guez-Gil}}{{Southworth} et~al.}{2009}]{2009AA...507..929S}
{Southworth} J.,  {Hickman} R.~D.~G.,  {Marsh} T.~R.,  {Rebassa-Mansergas} A.,
  {G{\"a}nsicke} B.~T.,  {Copperwheat} C.~M.,   {Rodr{\'\i}guez-Gil} P.,  2009,
  \mn@doi [\aap] {10.1051/0004-6361/200912885}, \href
  {https://ui.adsabs.harvard.edu/abs/2009A&A...507..929S} {507, 929}

\bibitem[\protect\citeauthoryear{{Spruit}}{{Spruit}}{1998}]{1998astro.ph..6141S}
{Spruit} H.~C.,  1998, arXiv e-prints, \href
  {https://ui.adsabs.harvard.edu/abs/1998astro.ph..6141S} {pp
  astro--ph/9806141}

\bibitem[\protect\citeauthoryear{{Steiner} \& {Diaz}}{{Steiner} \&
  {Diaz}}{1998}]{1998PASP..110..276S}
{Steiner} J.~E.,  {Diaz} M.~P.,  1998, \mn@doi [\pasp] {10.1086/316139}, \href
  {https://ui.adsabs.harvard.edu/abs/1998PASP..110..276S} {110, 276}

\bibitem[\protect\citeauthoryear{{Still}, {Dhillon}  \& {Jones}}{{Still}
  et~al.}{1995}]{1995MNRAS.273..849S}
{Still} M.~D.,  {Dhillon} V.~S.,   {Jones} D.~H.~P.,  1995, \mn@doi [\mnras]
  {10.1093/mnras/273.4.849}, \href
  {https://ui.adsabs.harvard.edu/abs/1995MNRAS.273..849S} {273, 849}

\bibitem[\protect\citeauthoryear{{Sytov}, {Kaigorodov}, {Bisikalo}, {Kuznetsov}
   \& {Boyarchuk}}{{Sytov} et~al.}{2007}]{2007ARep...51..836S}
{Sytov} A.~Y.,  {Kaigorodov} P.~V.,  {Bisikalo} D.~V.,  {Kuznetsov} O.~A.,
  {Boyarchuk} A.~A.,  2007, \mn@doi [Astronomy Reports]
  {10.1134/S1063772907100083}, \href
  {https://ui.adsabs.harvard.edu/abs/2007ARep...51..836S} {51, 836}

\bibitem[\protect\citeauthoryear{{Thoroughgood}, {Dhillon}, {Watson},
  {Buckley}, {Steeghs}  \& {Stevenson}}{{Thoroughgood}
  et~al.}{2004}]{2004MNRAS.353.1135T}
{Thoroughgood} T.~D.,  {Dhillon} V.~S.,  {Watson} C.~A.,  {Buckley} D.~A.~H.,
  {Steeghs} D.,   {Stevenson} M.~J.,  2004, \mn@doi [\mnras]
  {10.1111/j.1365-2966.2004.08135.x}, \href
  {https://ui.adsabs.harvard.edu/abs/2004MNRAS.353.1135T} {353, 1135}

\bibitem[\protect\citeauthoryear{{Thoroughgood} et~al.,}{{Thoroughgood}
  et~al.}{2005}]{2005MNRAS.357..881T}
{Thoroughgood} T.~D.,  et~al., 2005, \mn@doi [\mnras]
  {10.1111/j.1365-2966.2004.08613.x}, \href
  {https://ui.adsabs.harvard.edu/abs/2005MNRAS.357..881T} {357, 881}

\bibitem[\protect\citeauthoryear{{Tovmassian}, {Stephania Hernandez},
  {Gonz{\'a}lez-Buitrago}, {Zharikov}  \& {Garc{\'\i}a-D{\'\i}az}}{{Tovmassian}
  et~al.}{2014}]{2014AJ....147...68T}
{Tovmassian} G.,  {Stephania Hernandez} M.,  {Gonz{\'a}lez-Buitrago} D.,
  {Zharikov} S.,   {Garc{\'\i}a-D{\'\i}az} M.~T.,  2014, \mn@doi [\aj]
  {10.1088/0004-6256/147/3/68}, \href
  {https://ui.adsabs.harvard.edu/abs/2014AJ....147...68T} {147, 68}

\bibitem[\protect\citeauthoryear{{Walker}}{{Walker}}{1963}]{1963ApJ...137..485W}
{Walker} M.~F.,  1963, \mn@doi [\apj] {10.1086/147523}, \href
  {https://ui.adsabs.harvard.edu/abs/1963ApJ...137..485W} {137, 485}

\bibitem[\protect\citeauthoryear{{Wargau}, {Drechsel}, {Rahe}  \&
  {Bruch}}{{Wargau} et~al.}{1983}]{1983MNRAS.204P..35W}
{Wargau} W.,  {Drechsel} H.,  {Rahe} J.,   {Bruch} A.,  1983, \mn@doi [\mnras]
  {10.1093/mnras/204.1.35P}, \href
  {https://ui.adsabs.harvard.edu/abs/1983MNRAS.204P..35W} {204, 35P}

\bibitem[\protect\citeauthoryear{{Warner}}{{Warner}}{1973}]{1973IBVS..852....1W}
{Warner} B.,  1973, Information Bulletin on Variable Stars, \href
  {https://ui.adsabs.harvard.edu/abs/1973IBVS..852....1W} {852, 1}

\bibitem[\protect\citeauthoryear{{Warner}}{{Warner}}{1995}]{1995Ap&SS.226..187W}
{Warner} B.,  1995, \mn@doi [\apss] {10.1007/BF00627371}, \href
  {https://ui.adsabs.harvard.edu/abs/1995Ap&SS.226..187W} {226, 187}

\bibitem[\protect\citeauthoryear{{Williams}}{{Williams}}{1989}]{1989AJ.....97.1752W}
{Williams} R.~E.,  1989, \mn@doi [\aj] {10.1086/115115}, \href
  {https://ui.adsabs.harvard.edu/abs/1989AJ.....97.1752W} {97, 1752}

\bibitem[\protect\citeauthoryear{{Winkler}}{{Winkler}}{1977}]{1977AJ.....82.1008W}
{Winkler} L.,  1977, \mn@doi [\aj] {10.1086/112163}, \href
  {https://ui.adsabs.harvard.edu/abs/1977AJ.....82.1008W} {82, 1008}

\bibitem[\protect\citeauthoryear{{Woudt} \& {Warner}}{{Woudt} \&
  {Warner}}{2002}]{2002MNRAS.335...44W}
{Woudt} P.~A.,  {Warner} B.,  2002, \mn@doi [\mnras]
  {10.1046/j.1365-8711.2002.05613.x}, \href
  {https://ui.adsabs.harvard.edu/abs/2002MNRAS.335...44W} {335, 44}

\bibitem[\protect\citeauthoryear{{Young}, {Schneider}  \& {Shectman}}{{Young}
  et~al.}{1981}]{1981ApJ...244..259Y}
{Young} P.,  {Schneider} D.~P.,   {Shectman} S.~A.,  1981, \mn@doi [\apj]
  {10.1086/158703}, \href
  {https://ui.adsabs.harvard.edu/abs/1981ApJ...244..259Y} {244, 259}

\bibitem[\protect\citeauthoryear{{Zejda}, {Mikulasek}  \& {Wolf}}{{Zejda}
  et~al.}{2006}]{2006IBVS.5741....1Z}
{Zejda} M.,  {Mikulasek} Z.,   {Wolf} M.,  2006, Information Bulletin on
  Variable Stars, \href {https://ui.adsabs.harvard.edu/abs/2006IBVS.5741....1Z}
  {5741, 1}

\bibitem[\protect\citeauthoryear{{Zharikov}, {Tovmassian}, {Aviles}, {Michel},
  {Gonzalez-Buitrago}  \& {Garc{\'\i}a-D{\'\i}az}}{{Zharikov}
  et~al.}{2013}]{2013A&A...549A..77Z}
{Zharikov} S.,  {Tovmassian} G.,  {Aviles} A.,  {Michel} R.,
  {Gonzalez-Buitrago} D.,   {Garc{\'\i}a-D{\'\i}az} M.~T.,  2013, \mn@doi
  [\aap] {10.1051/0004-6361/201220099}, \href
  {https://ui.adsabs.harvard.edu/abs/2013A&A...549A..77Z} {549, A77}

\makeatother
\end{thebibliography}

\bsp	
\label{lastpage}
\end{document}